\documentclass[authoryear,final,1p,times]{elsarticle}
\usepackage{amsmath}    
\usepackage{amssymb}    
\usepackage{amsthm}     
\usepackage{amsfonts}
\usepackage[latin1]{inputenc}
\usepackage{latexsym}
\usepackage[english]{babel}
\usepackage[T1]{fontenc}
\usepackage{graphicx}
\usepackage{xspace}
\usepackage{multirow}
\journal{Arxiv}

\begin{document}

\newcommand{\pd}    {\backslash}
\newcommand{\Pn}    {\mathbb{P}}
\newcommand{\PP}    {{\bf P}}
\newcommand{\e}     {\ensuremath{\varepsilon}}
\newcommand{\dr}    {\ensuremath{\partial }}
\newcommand{\indep} {\ensuremath{\!\!\perp\!\!\!\perp\!\!}}
\newcommand{\transp}[1] {#1 ^{\top} }
\newcommand{\C}{\mathbb{C}}
\newcommand{\N}{\mathbb{N}}
\newcommand{\R}{\mathbb{R}}
\newcommand{\Z}{\mathbb{Z}}
\newcommand{\E}{\mathbb{E}}
\newcommand{\HH}{\mathbb{H}}
\newcommand{\K}{\mathbb{K}}
\newcommand{\T}{\mathbb{T}}
\newcommand{\cA}{{\mathcal A}}
\newcommand{\cB}{{\mathcal B}}
\newcommand{\cC}{{\mathcal C}}
\newcommand{\cD}{{\mathcal D}}
\newcommand{\cE}{{\mathcal E}}
\newcommand{\cF}{{\mathcal F}}
\newcommand{\cG}{{\mathcal G}}
\newcommand{\cH}{{\mathcal H}}
\newcommand{\cI}{{\mathcal I}}
\newcommand{\cJ}{{\mathcal J}}
\newcommand{\cK}{{\mathcal K}}
\newcommand{\cL}{{\mathcal L}}
\newcommand{\cM}{{\mathcal M}}
\newcommand{\cN}{{\mathcal N}}
\newcommand{\cO}{{\mathcal O}}
\newcommand{\cP}{{\mathcal P}}
\newcommand{\cR}{{\mathcal R}}
\newcommand{\cS}{{\mathcal S}}
\newcommand{\cT}{{\mathcal T}}
\newcommand{\cV}{{\mathcal V}}
\newcommand{\cU}{{\mathcal U}}
\newcommand{\cX}{{\mathcal X}}
\newcommand{\cY}{{\mathcal Y}}
\newcommand{\cZ}{{\mathcal Z}}
\def\bibname{References}
\def\refname{References}
\newcommand{\tend}  {\rightarrow}
\newcommand{\cvL}   {\stackrel{\it\mathcal Law}{\rightarrow} }
\newcommand{\cvP}   {\stackrel{\it\mathbb P}{\rightarrow} }
\newcommand{\cvps}  {\stackrel{\bf a.s.}{\rightarrow} }
\newtheorem{definition} {\bf Definition}[section]
\newtheorem{theoreme}   {\bf Theorem}[section]
\newtheorem{lemme}      {\bf lemme}[section]
\newtheorem{corollaire} {\bf Corollary}[section]
\newtheorem{proposition}{\bf Proposition}[section]
\newtheorem{PD}         {\bf Definition and Property}[section]
\newtheorem{PROPIE}     {\bf Property}[section]
\newtheorem{exemple}    {$\mathcal Example$}[section]
\newtheorem{remarque}   {$\mathcal Remark$ }[section]
\newtheorem{Sim}    {$\mathcal Simulation$}[section]
\newenvironment{preuve}        {\noindent {$\mathcal Proof\ :     $ \\}}{\hfill$\Box$\\}
\newenvironment{notation}      {\noindent {$\mathcal Notation\ :$ }\it}{}
\numberwithin{equation}{section}

\begin{frontmatter}
\title{PROJECTION PURSUIT THROUGH $\Phi$-DIVERGENCE MINIMISATION}
\author{Jacques Touboul}
\address{Université Pierre et Marie Curie\\
Laboratoire de Statistique Théorique et Appliquée\\
175 rue du Chevaleret, 75013 Paris, France\\
jack\_touboul@hotmail.com
}
\begin{abstract}
Consider a defined density on a set of very large dimension.  It is quite difficult to find an estimate of this density from a data set.  However, it is possible through a projection pursuit methodology to solve this problem.  In his seminal article, Huber (see "Projection pursuit", Annals of Statistics, 1985) demonstrates the interest of his method in a very simple given case. He considers the factorization of  density through a Gaussian component and some residual density. Huber's work is based on maximizing relative entropy.
Our proposal leads to a new algorithm.
Furthermore, we will also consider the case when the density to be factorized is estimated from an i.i.d. sample. We will then propose a test for the factorization of the estimated density.  Applications include a new test of fit pertaining to the Elliptical copulas.
\end{abstract}
\end{frontmatter}
\section{ Outline of the article}
The objective of Projection Pursuit is to generate one or several projections providing as much information as possible about the structure of the data set regardless of its size:

Once a structure has been isolated, the corresponding data are eliminated from the data set. Through a recursive approach, this process is iterated to find another structure in the remaining data, until no futher structure can be evidenced in the data left at the end.

\cite{Frie84} and \cite{MR790553} count among the first authors to have introduced this type of approaches for evidencing structures. They each describe, with many examples, how to evidence such a structure and consequently how to estimate the density of such data through two different methodologies each. Their work is based on maximizing relative entropy.\\
For a very long time, the two methodologies exposed by each of the above authors were thought to be equivalent but \cite{Zhu2004} showed it was in fact not the case when the number of iterations in the algorithms exceeds the dimension of the space containing the data. In the present article, we will therefore only focus on Huber's study while taking into account Mu Zhu remarks.

At present, let us briefly introduce Huber's methodology. We will then expose our approach and objective.
\subsection{ Huber's analytic approach}
Let $f$ be a density on $\R^d$. We define an instrumental density $g$ with same mean and variance as  $f$.
Huber's methodology requires us to start with performing the $K(f,g)=0$ test - with $K$ being the relative entropy. Should this test turn out to be positive, then $f=g$ and the algorithm stops. If the test were not to be verified, the first step of Huber's algorithm amounts to defining a vector $a_1$ and a density $f^{(1)}$  by 
\begin{equation}\label{DefSequMethodH1}
a_1\ =\ arg\inf_{a\in\R^d_*}\ K(f\frac{g_a}{f_a},g)\text{ and }f^{(1)}=f\frac{g_{a_1}}{f_{a_1}},
\end{equation}
where $\R^d_*$ is the set of non null vectors of $\R^d$, where $f_a$ (resp. $g_a$) stands for the density of $\transp aX$ (resp. $\transp aY$) when $f$ (resp. $g$) is the density of $X$ (resp. $Y$). More exactly, this results from the maximisation of $a\mapsto K(f_a,g_a)$ since $K(f,g)=K(f_a,g_a)+K(f\frac{g_a}{f_a},g)$ and it is assumed that $K(f,g)$ is finite.
In a second step, Huber replaces  $f$ with $f^{(1)}$ and goes through the first step again.\\
By iterating this process, Huber thus obtains a sequence $(a_1,a_2,...)$ of vectors of $\R^d_*$ and a sequence of densities $f^{(i)}$.
\begin{remarque}
Huber stops his algorithm when the relative entropy equals zero or when his algorithm reaches the $d^{th}$ iteration, he then obtains an approximation of $f$ from $g$ :\\
When there exists an integer $j$ such that $K(f^{(j)},g)=0$ with $j\leq d$, he obtains $f^{(j)}=g$, i.e. $f=g\Pi_{i=1}^j\frac{f^{(i-1)}_{a_i}} {g_{a_i}}$ since by induction $f^{(j)}=f\Pi_{i=1}^j\frac{g_{a_i}}{f^{(i-1)}_{a_i}}$. Similarly, when, for all $j$, Huber gets $K(f^{(j)},g)>0$ with $j\leq d$, he assumes $g=f^{(d)}$ in order to derive $f=g\Pi_{i=1}^d\frac{f^{(i-1)}_{a_i}} {g_{a_i}}$. \\
He can also stop his algorithm when the relative entropy equals zero without the condition $j\leq d$ is met. Therefore, since by induction we have $f^{(j)}=f\Pi_{i=1}^j\frac{g_{a_i}}{f^{(i-1)}_{a_i}}$ with $f^{(0)}=f$, we obtain $g=f\Pi_{i=1}^j\frac{g_{a_i}}{f^{(i-1)}_{a_i}}$. Consequently, we derive a representation of $f$ as $f=g\Pi_{i=1}^j\frac{f^{(i-1)}_{a_i}} {g_{a_i}}.$\\
Finally, he obtains  $K(f^{(0)},g)\geq K(f^{(1)},g)\geq.....\geq 0$ with $f^{(0)}=f$.
\end{remarque}
\subsection{ Huber's synthetic approach}
Keeping the notations of the above section, we start with performing the $K(f,g)=0$ test; should this test turn out to be positive, then $f=g$ and the algorithm stops, otherwise, the first step of his algorithm would consist in defining a vector $a_1$ and a density $g^{(1)}$  by 
\begin{equation}\label{DefSequMethodH2}
a_1\ =\ arg\inf_{a\in\R^d_*}\ K(f,g\frac{f_a}{g_a})\text{ and }g^{(1)}=g\frac{f_{a_1}}{g_{a_1}}.
\end{equation}
More exactly, this optimisation results from the maximisation of $a\mapsto K(f_a,g_a)$ since $K(f,g)=K(f_a,g_a)+K(f,g\frac{f_a}{g_a})$ and it is assumed that $K(f,g)$ is finite.
In a second step, Huber replaces  $g$ with $g^{(1)}$ and goes through the first step again.
By iterating this process, Huber thus obtains a sequence $(a_1,a_2,...)$ of vectors of $\R^d_*$ and a sequence of densities $g^{(i)}$.
\begin{remarque}
First, in a similar manner to the analytic approach, this methodology enables us to approximate and even to represent $f$ from $g$:\\
To obtain an approximation of $f$, Huber either stops his algorithm when the relative entropy equals zero, i.e. $K(f,g^{(j)})=0$ implies  $g^{(j)}=f$ with $j\leq d$, or when his  algorithm reaches the $d^{th}$ iteration, i.e. he approximates $f$ with $g^{(d)}$.\\
To obtain a representation of $f$, Huber stops his algorithm when the relative entropy equals zero, since $K(f,g^{(j)})=0$ implies  $g^{(j)}=f$. Therefore, since by induction we have $g^{(j)}=g\Pi_{i=1}^j\frac{f_{a_i}}{g^{(i-1)}_{a_i}}$ with $g^{(0)}=g$, we then obtain $f=g\Pi_{i=1}^j\frac{f_{a_i}}{g^{(i-1)}_{a_i}}.$\\
Second, he gets $K(f,g^{(0)})\geq K(f,g^{(1)})\geq.....\geq 0$ with $g^{(0)}=g$.
\end{remarque}
\subsection{Proposal}\label{OurProposal}
Let us first introduce the concept of $\Phi-$divergence. \\
Let $\varphi$ be a strictly convex function defined by $\varphi: \overline{\R^+}\tend\overline{\R^+},$ and such that $\varphi(1)=0$. We define a $\Phi-$divergence of $P$ from $Q$ - where $P$ and $Q$ are two probability distributions over a space $\Omega$ such that $Q$ is absolutely continuous with respect to $P$ - by  $$\Phi(Q,P)=\int\varphi(\frac{dQ}{dP})dP.$$
Throughout this article, we will also assume that $\varphi(0)<\infty$, that $\varphi'$ is continuous and that this divergence is greater than the $L^1$ distance - see also Annex \ref{FiDiv} page \pageref{FiDiv}.\\
Now, let us introduce our algorithm.
We start with performing the $\Phi(g,f)=0$ test; should this test turn out to be positive, then $f=g$ and the algorithm stops, otherwise, the first step of our algorithm would consist in defining a vector $a_1$ and a density $g^{(1)}$  by 
\begin{equation}\label{PhiDefSequMethod}
a_1\ =\ arg\inf_{a\in\R^d_*}\ \Phi(g\frac{f_a}{g_a},f)\text{ and }g^{(1)}=g\frac{f_{a_1}}{g_{a_1}}.
\end{equation}
Later on, we will prove that $a_1$ simultaneously optimises (\ref{DefSequMethodH1}), (\ref{DefSequMethodH2})  and (\ref{PhiDefSequMethod}). \\
In our second step, we will replace $g$ with $g^{(1)}$, and we will repeat the first step. \\And so on, by iterating this process, we will end up obtaining a sequence  $(a_1,a_2,...)$ of vectors in  $\R^d_*$  and a sequence of densities  $g^{(i)}$.
We will thus prove that the underlying structures of $f$ evidenced through this method are identical to the ones obtained through the  Huber's method. We will also evidence the above structures, which will enable us to infer more information on $f$ - see example below.
\begin{remarque}
As in the previous algorithm, we first provide an approximate and even a represention of $f$ from $g$:\\
To obtain an approximation of $f$, we stop our algorithm when the divergence equals zero, i.e. $\Phi(g^{(j)},f)=0$ implies  $g^{(j)}=f$ with $j\leq d$, or when our algorithm reaches the $d^{th}$ iteration, i.e. we approximate $f$ with $g^{(d)}$.\\
To obtain a representation of $f$, we stop our algorithm when the divergence equals zero. Therefore, since by induction we have $g^{(j)}=g\Pi_{i=1}^j\frac{f_{a_i}}{g^{(i-1)}_{a_i}}$ with $g^{(0)}=g$, we then obtain $f=g\Pi_{i=1}^j\frac{f_{a_i}}{g^{(i-1)}_{a_i}}.$\\
Second, he gets $\Phi(g^{(0)},f)\geq \Phi(g^{(1)},f)\geq.....\geq 0$ with $g^{(0)}=g$.\\
Finally, the specific form of relationship (\ref{PhiDefSequMethod}) establishes that we deal with M-estimation. We can therefore state that our method is more robust than Huber's - see \cite{BAR-YOHAI}, \cite{TOMA} as well as \cite{RobStat}.
\end{remarque}
\noindent At present, let us study two examples:
\begin{exemple}Let $f$ be a density defined on $\R^3$ by $f(x_1,x_2,x_3)=n(x_1,x_2)h(x_3)$, with $n$ being a bi-dimensional Gaussian density, and $h$ being a non Gaussian density. Let us also consider $g$, a Gaussian density with same mean and variance as $f$.\\
Since $g(x_1,x_2/x_3)=n(x_1,x_2)$, we then have $\Phi(g\frac{f_3}{g_3},f)=\Phi(n.f_3,f)=\Phi(f,f)=0$ as $f_3=h$, i.e. the function $a\mapsto \Phi(g\frac{f_a}{g_a},f)$ reaches zero for $e_3=(0,0,1)'$. \\ We therefore obtain $g(x_1,x_2/x_3)=f(x_1,x_2/x_3)$.
\end{exemple}
\begin{exemple}\label{exMarkov}
Assuming that the $\Phi$-divergence is greater than the $L^2$ norm.
Let us consider $(X_n)_{n\geq0}$, the Markov chain with continuous state space $E$.
Let $f$ be the density of $(X_0,X_1)$ and let $g$ be the normal density with same mean and variance as $f$.\\
Let us now assume that $\Phi(g^{(1)},f)=0$ with $g^{(1)}(x)=g(x)\frac{f_1}{g_1}$, i.e. let us assume that our algorithm stops for  $a_1=(1,0)'$. Consequently, if $(Y_0,Y_1)$ is a random vector with $g$ density, then the distribution law of $X_1$ given $X_0$ is Gaussian and is equal to the distribution law of $Y_1$ given $Y_0$.\\
And then, for any sequence $(A_i)$ - where $A_i\subset E$ - we have \\
$\PP\Big(X_{n+1}\in A_{n+1}\mid X_0\in A_{0}, X_1\in A_{1},\ldots, X_{n-1}\in A_{n-1},X_n\in A_{n}\Big)$

$\ \ \ \ \ \ \ \  \ \ $ $=\PP\left(X_{n+1}\in A_{n+1}\mid X_n\in A_{n}\right),$
based on  the very definition of a Markov chain,

$\ \ \ \ \ \ \ \  \ \ $ $=\PP\left(X_{1}\in A_{1}\mid X_0\in A_{0}\right),$
through the Markov property,

$\ \ \ \ \ \ \ \  \ \ $ $=\PP\left(Y_{1}\in A_{1}\mid Y_0\in A_{0}\right),$
as a consequence of the above nullity of the $\Phi$-divergence.
\end{exemple}
To recapitulate our method, if $\Phi(g,f)=0$, we derive $f$ from the relationship $f=g$; should a sequence $(a_i)_{i=1,...j}$, $j<d$, of vectors in $\R^d_*$ defining $g^{(j)}$ and such that $\Phi(g^{(j)},f)=0$ exist, then $f(./\transp{a_i}x, 1\leq i\leq j)=g(./\transp{a_i}x, 1\leq i\leq j)$, i.e. $f$ coincides with $g$ on the complement  of the vector subspace generated by the family $\{a_i\}_{i=1,...,j}$ - see also section 2 for a more detailed explanation. 

In this paper, after having clarified the choice of $g$,  we will consider the statistical solution to the representation problem, assuming that $f$ is unknown and $X_1$, $X_2$,... $X_m$ are i.i.d. with density $f$. We will provide asymptotic results pertaining to the family of optimizing vectors $a_{k,m}$ - that we will define more precisely below - as $m$ goes to infinity.
Our results also prove that the empirical representation scheme converges towards the theoretical one.
As an application, section \ref{copulaTest} permits a new test of fit pertaining to the copula of an unknown density $f$, section \ref{L1Conv} gives us an estimate of a density deconvoluted with a Gaussian component and section \ref{Regression} presents some applications to the regression analysis.
Finally, we will present simulations.
\section{ The algorithm}\label{TheAlgo}
\subsection{ The model}\label{modelSection}
As explained by \cite{Frie84} and \cite{MR0751274}, the choice of $g$ depends on the family of distribution one wants to find in $f$. Until now, the choice has only been to use the class of Gaussian distributions. This can be extended to the class of elliptic distributions with almost all $\Phi-$divergences.
\subsubsection{ Elliptical laws}
The interest of this class lies in the fact that conditional densities with elliptical distributions are also elliptical - see \cite{MR0629795}, \cite{MR2061237}. This very property allows us to use this class in our algorithm.
\begin{definition}
$X$ is said to abide by a multivariate elliptical distribution - noted $X\sim E_d(\mu,\Sigma,\xi_d)$ - if $X$ presents the following density, for any $x$ in $\R^d$ :

$\ \ \ \ \ \ \ \ \ \ \ \ \ \ \ \ \ \ \ \ \ \ \ \ \ \ \ \ $ $ f_X(x)=\frac{c_d}{|\Sigma|^{1/2}}\xi_d\Big(\frac{1}{2}(x-\mu)'\Sigma^{-1}(x-\mu)\Big)$\\
$\bullet$ with $\Sigma$, being a $d\times d$ positive-definite matrix and with $\mu$, being an $ d$-column vector,\\
$\bullet$ with $\xi_d$, being referred as the "density generator",\\
$\bullet$ with $c_d$, being a normalisation constant, such that $c_d=\frac{\Gamma(d/2)}{(2\pi)^{d/2}}\Big(\int_0^\infty x^{d/2-1}\xi_d(x)dx\Big)^{-1}$, \\with $\int_0^\infty x^{d/2-1}\xi_d(x)dx<\infty$.
\end{definition}
\begin{PROPIE}\label{ElliptProp}
1/ For any $X\sim E_d(\mu,\Sigma,\xi_d)$, for any $A$, being a $m\times d$ matrix with rank $m\leq d,$ and for any $b$, being an $m$-dimensional vector, we have $AX+b\sim E_m(A\mu+b,A\Sigma A',\xi_m)$.\\
Therefore, any marginal density of multivarite elliptical distribution is elliptic, i.e. \\
$X=(X_1,X_2,...,X_d)\sim E_d(\mu,\Sigma,\xi_d)\ \Rightarrow\ X_i\sim E_1(\mu_i,\sigma^2_i,\xi_1),$ $f_{X_i}(x)= \frac{c_1}{\sigma_i}\xi_1\Big(\frac{1}{2}(\frac{x-\mu_i}{\sigma})^2\Big),$ $1\leq i\leq d$.\\
2/ Corollary 5 of \cite{MR0629795} states that conditional densities with elliptical distributions are also elliptic. Indeed, if $X=(X_1,X_2)'\sim E_d(\mu,\Sigma,\xi_d)$, with $X_1$ (resp. $X_2$) being a size $d_1<d$ (resp. $d_2<d$), then $X_1/(X_2=a)\sim E_{d_1}(\mu',\Sigma',\xi_{d_1})$ with $\mu'=\mu_1+\Sigma_{12}\Sigma_{22}^{-1}(a-\mu_2)$ and $\Sigma'=\Sigma_{11}-\Sigma_{12}\Sigma_{22}^{-1}\Sigma_{21},$
with $\mu=(\mu_1,\mu_2)$ and $\Sigma=(\Sigma_{ij})_{1\leq i,j\leq 2}$.
\end{PROPIE}
\begin{remarque}\label{implyEstimBounded}
\cite{MR2061237} shows that multivariate Gaussian distributions derive from $\xi_d(x)=e^{-x}$. He also shows that if $X=(X_1,...,X_d)$ has an elliptical density such that its marginals verify $E(X_i)<\infty$ and $E(X_i^2)<\infty$ for $1\leq i\leq d,$ then $\mu$ is the mean of $X$ and $\Sigma$ is the covariance matrix of $X$. Consequently, from now on, we will assume that we are in this case.
\end{remarque}
\begin{definition}
Let $t$ be an elliptical density on $\R^k$ and let $q$ be an elliptical density on $\R^{k'}$.
The elliptical densities $t$ and $q$ are said to belong to the same family - or class - of elliptical densities, if their generating densities are $\xi_k$ and $\xi_{k'}$ respectively, which belong to a common given family of densities.
\end{definition}
\begin{exemple}
Consider two Gaussian densities  $\cN(0,1)$ and $\cN((0,0),Id_2)$. They are said to belong to the same elliptical families as they both present  $x\mapsto e^{-x}$ as generating density.
\end{exemple}
\subsubsection{ Choice of $g$}\label{gChoice}
Let us begin with studying the following case:\\
Let $f$ be a density on $\R^d$. Let us assume there exists $d$ not null independent vectors $a_j$, with $1\leq j\leq d,$ of $\R^d$, such that 
\begin{equation}\label{f-def}
f(x)=n(\transp{a_{j+1}}x,...,\transp{a_{d}}x)h(\transp{a_{1}}x,...,\transp{a_{j}}x),
\end{equation}
with $j<d$, with $n$ being an elliptical density on $\R^{d-j-1}$ and with $h$ being a density on $\R^{j}$, which does not belong to the same family as $n$. Let $X=(X_{1},...,X_{d})$ be a vector presenting $f$ as density.\\
Define $g$ as an Elliptical distribution with same mean and variance as $f$.\\
For simplicity, let us assume that the family $\{a_j\}_{1\leq j\leq d}$ is the canonical basis of $\R^d$:\\
The very definition of $f$ implies that $(X_{j+1},...,X_{d})$ is independent from $(X_{1},...,X_{j})$. Hence,  the density of $(X_{j+1},...,X_{d})$ given $(X_{1},...,X_{j})$ is $n$.\\
Let us assume that $\Phi(g^{(j)},f)=0,$ for some $j\leq d$. We then get $\frac{f(x)}{f_{a_1}f_{a_2}...f_{a_j}}=\frac{g(x)}{g^{(1-1)}_{a_1}g^{(2-1)}_{a_2}...g^{(j-1)}_{a_j}}$, since,  by induction, we have $g^{(j)}(x)=g(x)\frac{f_{a_1}}{g^{(1-1)}_{a_1}}\frac{f_{a_2}}{g^{(2-1)}_{a_2}}...\frac{f_{a_j}}{g^{(j-1)}_{a_j}}$.\\
Consequently, the fact that conditional densities with elliptical distributions are also elliptical as well as the above relationship enable us to infer that

$\ \ \ \ \ \ \  \ \ \ \ \ \ \  \ \ $    $n(\transp{a_{j+1}}x,.,\transp{a_{d}}x)=f(./\transp{a_i}x, 1\leq i\leq j)=g(./\transp{a_i}x, 1\leq i\leq j).$\\ In other words, $f$ coincides with $g$ on the complement  of the vector subspace generated by the family $\{a_i\}_{i=1,...,j}$.

Now, if the family $\{a_j\}_{1\leq j\leq d}$ is no longer the canonical basis of $\R^d$, then this family is again a basis of $\R^d$. Hence, lemma \ref{ChangBasis} - page \pageref{ChangBasis} - implies that
\begin{equation}\label{RelElli36}
g(./\transp{a_{1}}x,...,\transp{a_{j}}x)=n(\transp{a_{j+1}}x,...,\transp{a_{d}}x)=f(./\transp{a_{1}}x,...,\transp{a_{j}}x),
\end{equation}
which is equivalent to having $\Phi(g^{(j)},f)=0$ - since by induction  $g^{(j)}=g\frac{f_{a_1}}{g^{(1-1)}_{a_1}}\frac{f_{a_2}}{g^{(2-1)}_{a_2}}...\frac{f_{a_j}}{g^{(j-1)}_{a_j}}$.\\
The end of our algorithm implies that $f$ coincides with $g$ on the complement  of the vector subspace generated by the family $\{a_i\}_{i=1,...,j}$.
Therefore, the nullity of the $\Phi-$divergence provides us with information on the density structure.\\
\noindent In summary, the following proposition clarifies our choice of $g$ which depends on the family of distribution one wants to find in $f$ :
\begin{proposition}\label{Pb1}With the above notations, $\Phi(g^{(j)},f)=0$ is equivalent to
$$
g(./\transp{a_{1}}x,...,\transp{a_{j}}x)=f(./\transp{a_{1}}x,...,\transp{a_{j}}x)
$$
\end{proposition}
More generally, the above proposition leads us to defining the co-support of $f$ as the vector space generated from vectors $a_{1},...,a_{j}$.
\begin{definition}
Let $f$ be a density on $\R^d$. We define the co-vectors of $f$ as the sequence of vectors $a_{1},...,a_{j}$ which solves the problem $\Phi(g^{(j)},f)=0$ where $g$ is an Elliptical distribution with same mean and variance as $f$.
We define the co-support of $f$ as the vector space generated from vectors $a_{1},...,a_{j}$.
\end{definition}
\begin{remarque}Any $(a_i)$ family defining $f$ as in (\ref{f-def}), is an orthogonal basis of $\R^d$ - see lemma \ref{Base}
\end{remarque}
\subsection{Stochastic outline of our algorithm}\label{UseSample}
Let $X_1$, $X_2$,..,$X_m$ (resp. $Y_1$, $Y_2$,..,$Y_m$) be a sequence of $m$ independent random vectors with same density $f$ (resp. $g$). 
As customary in nonparametric $\Phi-$divergence optimizations, all estimates of $f$ and $f_a$ as well as all uses of Mont\'e Carlo's methods are being performed using subsamples $X_1$, $X_2$,..,$X_n$ and $Y_1$, $Y_2$,..,$Y_n$ - extracted respectively from $X_1$, $X_2$,..,$X_m$ and $Y_1$, $Y_2$,..,$Y_m$ - since the estimates are bounded below by some positive deterministic sequence $\theta_m$ - see Annex \ref{truncSample}.\\
Let $\Pn_n$ be the empirical measure of the subsample $X_1$, $X_2$,.,$X_n$. Let $f_n$ (resp.  $f_{a,n}$ for any $a$ in $\R^d_*$) be the kernel estimate of $f$ (resp.  $f_a$), which is built from $X_1$, $X_2$,..,$X_n$ (resp. $\transp aX_1$, $\transp aX_2$,..,$\transp aX_n$).\\
As defined in section \ref{OurProposal}, we introduce the following sequences $(a_{k})_{k\geq 1}$ and $(g^{(k)})_{k\geq 1}$:
\begin{eqnarray}\label{VraiDefOfAK}
&&\text{$\bullet$ $a_{k}$ is a non null vector of $\R^d$ such that $a_{k}=arg\min_{a\in\R^d_*} \Phi(g^{(k-1)}\frac{f_a}{g^{(k-1)}_a},f)$,}\\
&&\text{$\bullet$ $g^{(k)}$ is the density such that $g^{(k)}=g^{(k-1)}\frac{f_{a_{k}}}{g^{(k-1)}_{a_{k}}}$ with $g^{(0)}=g$.}\nonumber
\end{eqnarray}
The stochastic setting up of the algorithm uses $f_n$ and $g_n^{(0)}=g$ instead of $f$ and $g^{(0)}=g$ - since $g$ is known. Thus, at the first step, we build the vector $\check a_1$ which minimizes the $\Phi-$divergence between $f_n$ and $g\frac{f_{a,n}}{g_{a}}$ and which estimates $a_1$ :\\
Proposition \ref{QuotientDonneLoi} page \pageref{QuotientDonneLoi} and lemma \ref{toattain} page \pageref{toattain} enable us to minimize the $\Phi-$divergence between $f_n$ and $g\frac{f_{a,n}}{g_{a}}$. Defining $\check a_1$ as the argument of this minimization, proposition \ref{KernelpConv2} page \pageref{KernelpConv2} shows us that this vector tends to $a_1$.\\
Finally, we define the density $\check g^{(1)}_m$ as $\check g^{(1)}_m=g\frac{f_{\check a_1,m}}{g_{\check a_1}}$ which estimates $g^{(1)}$ through theorem \ref{KernelKRessultatPricipal}.\\
Now, from the second step and as defined in section \ref{OurProposal}, the density $g^{(k-1)}$ is unknown. Consequently, once again, we have to truncate the samples:\\
All estimates of $f$ and $f_a$ (resp. $g^{(1)}$ and $g_a^{(1)}$) are being performed using a subsample $X_1$, $X_2$,..,$X_n$ (resp. $Y_1^{(1)}$, $Y_2^{(1)}$,..,$Y_n^{(1)}$) extracted from $X_1$, $X_2$,..,$X_m$ (resp. $Y_1^{(1)}$, $Y_2^{(1)}$,..,$Y_m^{(1)}$ - which is a sequence of $m$ independent random vectors with same density $g^{(1)}$) such that the estimates are bounded below by some positive deterministic sequence $\theta_m$ - see Annex \ref{truncSample}.\\
Let $\Pn_n$ be the empirical measure of the subsample $X_1$, $X_2$,..,$X_n$. Let $f_n$ (resp. $g_n^{(1)}$, $f_{a,n}$, $g_{a,n}^{(1)}$ for any $a$ in $\R^d_*$) be the kernel estimate of $f$ (resp. $g^{(1)}$ and $f_a$ as well as $g_a^{(1)}$) which is built from $X_1$, $X_2$,..,$X_n$ (resp. $Y_1^{(1)}$, $Y_2^{(1)}$,..,$Y_n^{(1)}$ and $\transp aX_1$, $\transp aX_2$,..,$\transp aX_n$ as well as $\transp aY_1^{(1)}$, $\transp aY^{(1)}_2$,..,$\transp aY_n^{(1)}$).
The stochastic setting up of the algorithm uses $f_n$ and $g_n^{(1)}$ instead of $f$ and $g^{(1)}$.
Thus, we build the vector $\check a_2$ which minimizes the $\Phi-$divergence between $f_n$ and $g_n^{(1)}\frac{f_{a,n}}{g_{a,n}^{(1)}}$ - since $g^{(1)}$ and $g^{(1)}_a$ are unknown - and which estimates $a_2$.
Proposition \ref{QuotientDonneLoi} page \pageref{QuotientDonneLoi} and lemma \ref{toattain} page \pageref{toattain} enable us to minimize the $\Phi-$divergence between $f_n$ and $g_n^{(1)}\frac{f_{a,n}}{g_{a,n}^{(1)}}$. Defining $\check a_2$ as the argument of this minimization, proposition \ref{KernelpConv2} page \pageref{KernelpConv2} shows us that this vector tends to $a_2$ in $n$. Finally, we define the density $\check g^{(2)}_n$ as $\check g^{(2)}_n=g_n^{(1)}\frac{f_{\check a_2,n}}{g^{(1)}_{\check a_2,n}}$ which estimates $g^{(2)}$ through theorem \ref{KernelKRessultatPricipal}.\\
And so on, we will end up obtaining a sequence  $(\check a_1,\check a_2,...)$ of vectors in  $\R^d_*$ estimating the co-vectors of $f$ and a sequence of densities $(\check g^{(k)}_n)_k$ such that $\check g^{(k)}_n$ estimates $g^{(k)}$ through theorem \ref{KernelKRessultatPricipal}.
\section{Results}\label{OurResult}
\subsection{ Convergence results}\label{cvResult}
\subsubsection{ Hypotheses on $f$}\label{HypoF}
In this paragraph, we define the set of hypotheses on $f$ which could possibly be of use in our work. Discussion on several of these hypotheses can be found in Annex \ref{DiscussHyp}.\\
In this section, to be more legible we replace $g$ with $g^{(k-1)}$. Let

$\ \ \ \ \ \ \  \ \ \ \ \ \ \ $ $\Theta =\R^d,\ \Theta^{\Phi} =\{b\in\Theta\ |\ \ \int\varphi^*(\varphi'(\frac{g(x)}{f(x)}\frac{f_b(\transp bx)}{g_b(\transp bx)}))d{\bf P}<\infty\},$

$\ \ \ \ \ \ \  \ \ \ \ \ \ \ $ $M(b,a,x)=\int\varphi'(\frac{g(x)}{f(x)}\frac{f_b(\transp bx)}{g_b(\transp bx)})g(x)\frac{f_a(\transp ax)}{g_a(\transp ax)}dx-\ \varphi^*(\varphi'(\frac{g(x)}{f(x)}\frac{f_b(\transp bx)}{g_b(\transp bx)})),$

$\ \ \ \ \ \ \  \ \ \ \ \ \ \ $ $\Pn_n M(b,a)=\int M(b,a,x)d\Pn_n,$ ${\bf P} M(b,a)=\int M(b,a,x) d{\bf P},$\\
where $\PP $ is the probability measure presenting $f$ as density.\\
Similarly as in chapter $V$ of \cite{MR1652247}, let us define :\\
$(H1)$ : $\text{For all $\e>0$, there is $\eta>0$, such that for all $c\in\Theta^\Phi $ verifying }\|c-a_k\|\geq \e,$

$\ \ \ $ $\text{ we have }{\bf P} M(c,a)-\eta>{\bf P} M(a_k,a),\text{ with }a\in\Theta.$\\
$(H2)$ : $\text{$\exists$ $Z<0$, $n_0>0$ such that } (n\geq n_0\ \Rightarrow \ \sup_{a\in\Theta}\sup_{c\in\{\Theta^\Phi \}^c}\Pn_nM(c,a)<Z)$\\
$(H3)$ : $\text{There is a neighbourhood $V$ of $a_k$, and a positive function $H$, such that,}$

$\ \ \ $ $\text{ for all $c\in V$, we have }|M(c,a_k,x)|\leq H(x)\ ({\bf P} -a.s.)\text{ with }{\bf P} H<\infty,$\\
$(H4)$ : $\text{There is a neighbourhood $V$ of $a_k$, such that for all $\e$, there is a $\eta$ such that for}$

$\ \ \ $ $\text{ all $c \in V$ and $a\in\Theta$, verifying }\|a-a_k\|\geq \e,\text{ we have }{\bf P} M(c,a_k)<{\bf P} M(c,a)-\eta.$\\
Putting $I_{a_k}=\frac{\dr^2}{\dr a^2}   \Phi (g\frac{f_{a_k}}{g_{a_k}},f),$
and $x\tend \rho(b,a,x)=\varphi'(\frac{g(x)f_b(\transp bx)}{f(x)g_b(\transp bx)})\frac{g(x)f_a(\transp ax)}{g_a(\transp ax)}$, let us now consider three new hypotheses:\\
$(H5)$ : The function $\varphi$ is ${\mathcal C}^3$ in $(0,+\infty$) and there is a neighbourhood $V_k'$ of $(a_k,a_k)$ such that, for 

$\ \ \ \ $ all  $(b,a)$ of $V_k'$, the gradient $\nabla (\frac{g(x)f_a(\transp ax)}{g_a(\transp ax)})$ and the Hessian ${\mathcal H}(\frac{g(x)f_a(\transp ax)}{g_a(\transp ax)})$ exist ($\lambda \_a.s.$), and

$\ \ \ $ the first order partial derivatives $\frac{g(x)f_a(\transp ax)}{g_a(\transp ax)}$ and the first and second order derivatives  of  

$\ \ \ $ $(b,a)\mapsto \rho(b,a,x)$ are dominated  ($\lambda\_$a.s.) by $\lambda$-integrable functions.\\
$(H6)$ : The function $(b,a)\mapsto M(b,a)$ is ${\mathcal C}^3$ in a neighbourhood $V_k$ of $(a_k,a_k)$ for all $x$; and the 

$\ \ \ $  partial derivatives of $(b,a)\mapsto M(b,a)$ are all dominated in $V_k$ by a ${\bf P}\_$integrable  function  

$\ \ \ $ $H(x)$.\\
$(H7)$ : ${\bf P}\|\frac{\dr}{\dr b}M(a_k,a_k)\|^2$ and ${\bf P}\|\frac{\dr}{\dr a}M(a_k,a_k)\|^2$ are finite and the expressions ${\bf P}\frac{\dr^2}{\dr b_i\dr b_j}M(a_k,a_k)$ and  

$\ \ \ $ $I_{a_k}$ exist and are invertible.\\
Finally, we define\\
$(H8)$ : There exists $k$ such that  ${\bf P} M(a_k,a_k)= 0$.\\
$(H9)$ : $(Var_{{\bf P}}(M(a_k,a_k)))^{1/2}$ exists and is invertible.\\
$(H0)$: $f$ and $g$ are assumed to be positive and bounded.
\subsubsection{ Estimation of the first co-vector of $f$}\label{Estimofa1}
Let $\cR$ be the class of all positive functions $r$ defined on $\R$ and such that $g(x)r(\transp ax)$ is a density on $\R^d$ for all $a$ belonging to $\R^d_*$. The following proposition shows that there exists a vector $a$ such that $\frac{f_a}{g_a}$ minimizes $\Phi(gr,f)$ in $r$:
\begin{proposition} \label{lemmeHuberModifprop}
There exists a vector $a$ belonging to $\R^d_*$ such that
$$
arg\min_{r\in\cR}\Phi(gr,f)=\frac{f_a}{g_a}\text{ and }r(\transp ax)=\frac{f_a(\transp ax)}{g_a(\transp ax)}.
$$
\end{proposition}
\begin{remarque}\label{criteria}
This proposition proves that $a_1$ simultaneously optimises (\ref{DefSequMethodH1}), (\ref{DefSequMethodH2}) and (\ref{PhiDefSequMethod}). In other words, it proves that the underlying structures of $f$ evidenced through our method are identical to the ones obtained through Huber's methods - see also Annex \ref{a1solve4}.
\end{remarque}
Following \cite{MR2054155}, let us introduce the estimate of $\Phi(g\frac{f_{a,n}}{g_{a}},f_{n})$, through $$\check   \Phi(g\frac{f_{a,n}}{g_{a}},f_{n})=\int M(a,a,x) d\Pn_n(x)$$
\begin{proposition}\label{info}
Let $\check  a$ be such that $\check  a   :=  arg\inf_{a\in\R^d_*}\check   \Phi(g\frac{f_{a,n}}{g_{a}},f_{n}).$\\
Then, $\check  a$ is a strongly convergent estimate of $a$, as defined in proposition \ref{lemmeHuberModifprop}.
\end{proposition}
\noindent Let us also introduce the following sequences  $(\check a_{k})_{k\geq 1}$ and $(\check g^{(k)}_{n})_{k\geq 1}$, for any given $n$ - see section \ref{UseSample}.:\\
$\bullet$ $\check a_{k}$ is an estimate of $a_{k}$ as defined in proposition \ref{info}  with $\check g^{(k-1)}_{n}$ instead of $g$,\\
$\bullet$ $\check g^{(k)}_{n}$ is such that $\check g^{(0)}_{n}=g$, $\check g^{(k)}_{n}(x)=\check g^{(k-1)}_{n}(x)\frac{f_{\check a_k,n}(\transp {\check a_k}x)}{[\check g^{(k-1)}]_{\check a_k,n}(\transp {\check a_k}x)}$, i.e.
$\check g^{(k)}_{n}(x)=g(x)\Pi_{j=1}^k\frac{f_{\check a_j,n}(\transp {\check a_j}x)}{[\check g^{(j-1)}]_{\check a_j,n}(\transp {\check a_j}x)}$.\\
We also note that $\check g^{(k)}_n$ is a density.
\subsubsection{ Convergence study at the  $k^{\text{th}}$ step of the algorithm:}
In this paragraph, we will show that the sequence $(\check a_k)_n$ converges towards $a_k$ and that the sequence $(\check g^{(k)}_n)_n$ converges towards $g^{(k)}$.\\ 
Let $\check c_n(a)    =\ arg\sup_{c\in\Theta }\ \Pn_nM(c,a),$ with $a\in\Theta$,
and $\check \gamma_n   =\ arg\inf_{a\in\Theta }\ \sup_{c\in\Theta }\ \Pn_nM(c,a)$.
We state
\begin{proposition}\label{KernelpConv2}	Both $\sup_{a\in\Theta}\|\check c_n(a)-a_k\|$  and  $\check \gamma_n $ converge toward $a_k$ a.s.
\end{proposition}
\noindent Finally, the following theorem shows that $\check  g^{(k)}_n$ converges almost everywhere towards $g^{(k)}$:
\begin{theoreme}\label{KernelKRessultatPricipal}It holds
$\check  g^{(k)}_n\to_n   g^{(k)}\ a.s.$
\end{theoreme}
\subsection{ Asymptotic Inference at the  $k^{\text{th}}$ step of the algorithm}\label{AsymptResult}
The following theorem shows that $\check g^{(k)}_n$ converges towards $ g^{(k)}$ at the rate $O_{\PP}(n^{-\frac{2}{2+d}})$ in three differents cases, namely for any given $x$, with the  $L^1$ distance and with the relative  entropy:
\begin{theoreme}\label{KernelSuperdiffconv}It holds 
$|\check g^{(k)}_n(x)-g^{(k)}(x)|=O_{\PP}(n^{-\frac{2}{2+d}}),$
$\int |\check g^{(k)}_n(x)-g^{(k)}(x)|dx=O_{\PP}(n^{-\frac{2}{2+d}})$ and
$|K(\check g^{(k)}_n,f)-K(g^{(k)},f)|=O_{\PP}(n^{-\frac{2}{2+d}}).$
\end{theoreme}
\begin{remarque}
With the relative entropy, we have $n=O(m^{1/2})$ - see lemma \ref{n-FunctOf-m}. The above rates consequently become $O_{\PP}(m^{-\frac{1}{2+d}})$.
\end{remarque}
The following theorem shows that the laws of our estimators of $a_k$, namely $\check c_n(a_k)$ and $\check \gamma_n$, converge towards a linear combination of Gaussian variables.
\begin{theoreme}\label{Kernelfiloiestimateurs} 
It holds\\
$\sqrt n{\mathcal A}.(\check c_n(a_k)-a_k)\cvL {\mathcal B}.{\mathcal N}_d(0,{\bf P}\|\frac{\dr}{\dr b}M(a_k,a_k)\|^2)+{\mathcal C}.{\mathcal N}_d(0,{\bf P}\|\frac{\dr}{\dr a}M(a_k,a_k)\|^2)$ and\\
$\sqrt n{\mathcal A}.(\check \gamma_n-a_k)\cvL {\mathcal C}.{\mathcal N}_d(0,{\bf P}\|\frac{\dr}{\dr b}M(a_k,a_k)\|^2)+{\mathcal C}.{\mathcal N}_d(0,{\bf P}\|\frac{\dr}{\dr a}M(a_k,a_k)\|^2)$\\
where ${\mathcal A}={\bf P}\frac{\dr^2}{\dr b\dr b}M(a_k,a_k)(\PP\frac{\dr^2}{\dr a_i\dr a_j}M(a_k,a_k)+\PP\frac{\dr^2}{\dr a_i\dr b_j}M(a_k,a_k))$, 
${\mathcal C}={\bf P}\frac{\dr^2}{\dr b\dr b}M(a_k,a_k)$ and\\
${\mathcal B}={\bf P}\frac{\dr^2}{\dr b\dr b}M(a_k,a_k)+\PP\frac{\dr^2}{\dr a_i\dr a_j}M(a_k,a_k)+\PP\frac{\dr^2}{\dr a_i\dr b_j}M(a_k,a_k).$
\end{theoreme}
\subsection{A stopping rule for the procedure}\label{StopAlgo}
In this paragraph, we will call $\check g^{(k)}_n$ (resp. $\check g^{(k)}_{a,n}$) the kernel estimator of  $\check g^{(k)}$ (resp. $\check g^{(k)}_{a}$). We will first show that $g_n^{(k)}$ converges towards $f$ in $k$ and $n$. Then, we will provide a stopping rule for this identification procedure.
\subsubsection{ Estimation of $f$}\label{EstOfF}
The following proposition provides us with an estimate of $f$:
\begin{theoreme}\label{limnk}
We have $\lim_n\lim_k\check g^{(k)}_n=f$ a.s.
\end{theoreme}
Consequently, the following corollary shows that $\Phi(g^{(k-1)}_n\frac{f_{a_k,n}}{g^{(k-1)}_{a_k,n}},f_{a_k,n})$ converges towards zero as $k$ and then as $n$ go to infinity: 
\begin{corollaire}\label{FromSection14Huber-1}
We have $\lim_n\lim_k\Phi(\check g^{(k)}_n\frac{f_{{a_k},n}}{[\check g^{(k)}]_{{a_k},n}},f_n)= 0$ a.s.
\end{corollaire}
\subsubsection{ Testing of the criteria}\label{Test}
In this paragraph, through a test of our criteria,  namely $a\mapsto \Phi(\check g^{(k)}_n\frac{f_{a,n}}{[\check g^{(k)}]_{a,n}},f_n)$, we will build a stopping rule for this procedure.\\
First, the next theorem enables us to derive the law of our criteria:
\begin{theoreme} \label{KernelLOIDUCRITERE} For a fixed $k$, we have 

$\sqrt n(Var_{\PP}(M(\check c_n(\check \gamma_n),\check \gamma_n)))^{-1/2}(\Pn_nM(\check c_n(\check \gamma_n),\check \gamma_n)-\Pn_nM(a_k,a_k)) \cvL \cN(0,I)$,\\ where $k$ represents the $k^{th}$ step of our algorithm  and where $I$ is the identity matrix in $\R^d$.
\end{theoreme}
Note that $k$ is fixed in theorem \ref{KernelLOIDUCRITERE} since $\check \gamma_n   =\ arg\inf_{a\in\Theta }\ \sup_{c\in\Theta }\ \Pn_nM(c,a)$ where $M$ is a known function of $k$ - see section \ref{HypoF}. Thus, in the case when $\Phi(g^{(k-1)}\frac{f_{a_k}}{g^{(k-1)}_{a_k}},f)= 0$, we obtain
\begin{corollaire} \label{KernelLOIDUCRITERE2} 
We have $\sqrt n(Var_{\PP}(M(\check c_n(\check \gamma_n),\check \gamma_n)))^{-1/2}\Pn_nM(\check c_n(\check \gamma_n),\check \gamma_n) \cvL \cN(0,I)$.
\end{corollaire} 
\noindent Hence, we propose the test of the null hypothesis $$\text{$(H_0)$ : $\Phi(g^{(k-1)}\frac{f_{a_k}}{g^{(k-1)}_{a_k}},f)= 0$ versus the alternative $(H_1)$ :  $\Phi(g^{(k-1)}\frac{f_{a_k}}{g^{(k-1)}_{a_k}},f)\not= 0$.}$$
Based on this result, we stop the algorithm, then, defining $a_k$ as the last vector generated, we derive from corollary \ref{KernelLOIDUCRITERE2} a $\alpha$-level confidence ellipsoid around $a_k$, namely $$\cE_k=\{b\in\R^d;\ \sqrt n(Var_{\PP}(M(b,b)))^{-1/2}\Pn_nM(b,b)\leq q_{\alpha}^{\cN(0,1)} \}$$
where $q_{\alpha}^{\cN(0,1)}$ is the quantile of a $\alpha$-level reduced centered normal distribution and where $\Pn_n$ is the empirical measure araising from a realization of the sequences $(X_1,\ldots,X_n)$ and $(Y_1,\ldots,Y_n)$.\\ Consequently, the following corollary provides us with a confidence region for the above test:
\begin{corollaire}\label{KernelLOIDUCRITERE2coro}
$\cE_k$ is a confidence region for the test of the null hypothesis $(H_0)$ versus $(H_1)$.
\end{corollaire}
\subsection{ Goodness-of-fit test for copulas}\label{copulaTest}
Let us begin with studying the following case:\\
Let $f$ be a density defined on $\R^2$ and let $g$ be an Elliptical distribution with same mean and variance as $f$.
Assuming first that our algorithm leads us to having $\Phi(g^{(2)},f)=0$ where family $(a_i)$ is the canonical basis of $\R^2$. Hence,  we have $g^{(2)}(x)=g(x)\frac{f_1}{g_1}\frac{f_2}{g^{(1)}_2}=g(x)\frac{f_1}{g_1}\frac{f_2}{g_2}$ - through lemma \ref{TrucBidule} page \pageref{TrucBidule} - and $g^{(2)}=f$.
Therefore, $f=g(x)\frac{f_1}{g_1}\frac{f_2}{g_2},$ i.e. $\frac{f}{f_1f_2}=\frac{g}{g_1g_2}$, and then $$\frac{\dr^2}{\dr x\dr y}C_f=\frac{\dr^2}{\dr x\dr y}C_g$$ where $C_f$ (resp. $C_g$) is the copula of $f$ (resp. $g$).\\
At present, let $f$ be a density on $\R^d$ and let $g$ be the density defined in section \ref{gChoice}.\\
Let us assume that our algorithm implies that $\Phi(g^{(d)},f)=0$.\\
Hence, we have, for any $x\in\R^d$, $ g(x)\Pi_{k=1}^d\frac{f_{a_k}(\transp {a_k}x)}{[g^{(k-1)}]_{a_k}(\transp {a_k}x)}=f(x)$, i.e. $
\frac{g(x)}{\Pi_{k=1}^d g_{a_k}(\transp {a_k}x)}=\frac{f(x)}{\Pi_{k=1}^df_{a_k}(\transp {a_k}x)}$,
since lemma \ref{TrucBidule} page \pageref{TrucBidule} implies that $g^{(k-1)}_{a_k}=g_{a_k}$ if $k\leq d$.\\
Moreover, the family $(a_i)_{i=1...d}$ is a basis of $\R^d$ - see lemma \ref{imFree} page \pageref{imFree}. Hence, putting $A=(a_1,...,a_d)$ and defining vector $y$ (resp. density $\tilde f$, copula $\tilde C_f$ of $\tilde f$, density $\tilde g$, copula $\tilde C_g$ of $\tilde g$) as the expression of vector $x$ (resp. density $f$, copula $C_f$ of $ f$, density $g$, copula $C_g$ of $g$) in basis $A$, the above equality implies
$$\frac{\dr^d}{\dr y_1...\dr y_d}\tilde C_f=\frac{\dr^d}{\dr y_1...\dr y_d}\tilde C_g.$$
Finally, we perform a statistical test of the null hypothesis $(H_0)$ : $\frac{\dr^d}{\dr y_1...\dr y_d}\tilde C_f=\frac{\dr^d}{\dr y_1...\dr y_d}\tilde C_g$ versus the alternative $(H_1)$ : $\frac{\dr^d}{\dr y_1...\dr y_d}\tilde C_f\not=\frac{\dr^d}{\dr y_1...\dr y_d}\tilde C_g$. Since, under $(H_0)$, we have $\Phi(g^{(d)},f)=0$, then, as explained in section \ref{Test},  corollary \ref{KernelLOIDUCRITERE2coro} provides us with a confidence region for our test.
\begin{theoreme}\label{TestCop}
Keeping the notations of corollary \ref{KernelLOIDUCRITERE2coro}, we infer that
$\cE_d$ is a confidence region for the test of the null hypothesis $(H_0)$ versus the alternative hypothesis $(H_1)$.\end{theoreme}
\subsection{Rewriting of the convolution product}\label{L1Conv}
In the present paper, we first elaborated an algorithm aiming at isolating several known structures from initial datas. Our objective was to verify if for a known density on $\R^d$,  a known density $n$ on $\R^{d-j-1}$ such that, for $d>1$, 
\begin{equation}\label{f-def-rap}
f(x)=n(\transp{a_{j+1}}x,...,\transp{a_{d}}x)h(\transp{a_{1}}x,...,\transp{a_{j}}x),
\end{equation}
did indeed exist, with $j<d$, with $(a_1,\ldots,a_d)$ being a basis of $\R^d$ and with $h$ being a density on $\R^{j}$.\\
Secondly, our next step consisted in building an estimate (resp. a representation) of $f$ without necessarily assuming that $f$ meets relationship (\ref{f-def-rap}) - see theorem \ref{limnk}.\\
Consequently, let us consider $Z_1$ and $Z_2$, two random vectors with respective densities $h_1$ and $h_2$ - which is Elliptical - on $\R^d$. Let us consider a random vector $X$ such that $X=Z_1+Z_2$ and let $f$ be its density. This density can then be written as :
$$
f(x)=h_1*h_2(x)=\int_{\R^d}h_1(x)h_2(t-x)dt.
$$
Then, the following property enables us to represent $f$ under the form of a product and without the integral sign
\begin{proposition}\label{HuberApp}
Let $\phi$ be a centered Elliptical density with  $\sigma^2.I_d$, $\sigma^2>0$, as covariance matrix, such that it is a product density in all orthogonal coordinate systems and such that its characteristic function $s\mapsto \Psi(\frac{1}{2}|s|^2\sigma^2)$ is integrable - see \cite{MR2061237}.\\
Let  $f$ be a density on $\R^d$ which can be deconvoluted with $\phi$, i.e. 
$$f=\overline f*\phi=\int_{\R^d}\overline f(x)\phi(t-x)dt,$$
where $\overline f$ is some density on $\R^d$.\\
Let $g^{(0)}$ be the Elliptical density belonging to the same Elliptical family as $f$ and having same mean and variance as $f$. \\
Then, the sequence $(g^{(k)})_k$ converges uniformly a.s. and in $L^1$ towards $f$ in $k$, i.e. 
$$
\lim_{k\to\infty}\sup_{x\in\R^d}|g^{(k)}(x)-f(x)|=0,\text{ and }\lim_{k\to\infty}\int_{\R^d}|g^{(k)}(x)-f(x)|dx=0.
$$
\end{proposition}
\noindent Finally, with the notations of section \ref{StopAlgo} and of proposition \ref{HuberApp}, the following theorem enables us to estimate any convolution product of a multivariate Elliptical density $\phi$ with a continuous density $\overline f$:
\begin{theoreme}\label{Sum=Product}
It holds $\lim_n\lim_k\check g^{(k)}_n=\overline f*\phi$ $a.s.$
\end{theoreme}
\subsection{On the regression}\label{Regression}

In this section, we will study several applications of our algorithm pertaining to the regression analysis. We define $(X_1,...,X_d)$ (resp. $(Y_1,...,Y_d)$) as a vector with density $f$  (resp. $g$ - see section \ref{gChoice}).
\begin{remarque}
In this paragraph, we will work in the $L^2$ space. Then, we will first only consider the $\Phi-$divergences which are greater than or equal to the $L^2$ distance - see \cite{MR0356302}. Note also that the co-vectors of $f$ can be obtained in the $L^2$ space - see lemma \ref{toattain} and proposition \ref{QuotientDonneLoi}.
\end{remarque}
\subsubsection{The basic idea}\label{MarkovSection}
In this paragraph, we will assume that $\Theta=\R^2_*$ and that our algorithm stops for $j=1$ and $a_1=(0,1)'$.
The following theorem provides us with the regression of $X_1$ on $X_2$ :
\begin{theoreme}\label{FirstReg}
The probability measure of $X_1$ given $X_2$ is the same as the probability measure of $Y_1$ given $Y_2$. Moreover, the regression between $X_1$ and $X_2$ is
$$
X_1=E(Y_1/Y_2)+\varepsilon,
$$
where $\varepsilon$ is a centered random variable orthogonal to $E(X_1/X_2)$.\\
\end{theoreme}
\begin{remarque}\label{FirstRegRem}
This theorem implies that $E(X_1/X_2)=E(Y_1/Y_2)$. This equation can be used in many fields of research. The Markov chain theory has been used for instance in example \ref{exMarkov}.\\
Moreover, if $g$ is a Gaussian density with same mean and variance as $f$, then \cite{Sap06} implies that
$E(Y_1/Y_2)=E(Y_1)+\frac{Cov(Y_1,Y_2)}{Var(Y_2)}(Y_2-E(Y_2))$ and then
$$
X_1=E(Y_1)+\frac{Cov(Y_1,Y_2)}{Var(Y_2)}(Y_2-E(Y_2))+\varepsilon.
$$
\end{remarque}
\subsubsection{General case}\label{RegGenCase}
In this paragraph, we will assume that $\Theta=\R^d_*$ and that our algorithm stops with $j$ for $j<d$.
Lemma \ref{OrthoOfVect} implies the existence of an orthogonal and free family $(b_i)_{i=j+1,..,d}$ of $\R^d_*$ such that $\R^d=Vect\{a_i\}\stackrel{\perp}{\oplus}Vect\{b_k\}$ and such that
\begin{eqnarray}\label{RegGenRel}
g(\transp{b_{j+1}}x,...,\transp{b_{d}}x/\transp{a_{1}}x,...,\transp{a_{j}}x)=f(\transp{b_{j+1}}x,...,\transp{b_{d}}x/\transp{a_{1}}x,...,\transp{a_{j}}x).
\end{eqnarray}
Hence, the following theorem provides us with the regression of $\transp{b_{k}}X$, $k=1,...,d$, on $(\transp{a_{1}}X,...,\transp{a_{j}}X)$:
\begin{theoreme}\label{SecondReg}
The probability measure of $(\transp{b_{j+1}}X,...,\transp{b_{d}}X)$ given $(\transp{a_{1}}X,...,\transp{a_{j}}X)$ is the same as the probability measure of $(\transp{b_{j+1}}Y,...,\transp{b_{d}}Y)$ given $(\transp{a_{1}}Y,...,\transp{a_{j}}Y)$. Moreover, the regression of $\transp{b_{k}}X$, $k=1,...,d$, on $(\transp{a_{1}}X,...,\transp{a_{j}}X)$ is
$\transp{b_{k}}X=E(\transp{b_{k}}Y/\transp{a_{1}}Y_1,...,\transp{a_{j}}Y)+\transp{b_{k}}\varepsilon$, where $\varepsilon$ is a centered random vector such that $\transp{b_{k}}\varepsilon$ is orthogonal to $E(\transp{b_{k}}X/\transp{a_{1}}X,...,\transp{a_{j}}X)$.
\end{theoreme}
\begin{corollaire}\label{SecondRegCoro}
If $g$ is a Gaussian density with same mean and variance as $f$, and if $Cov(X_i,X_j)=0$ for any $i\not=j$, then,
the regression of $\transp{b_{k}}X$, $k=1,...,d$, on $(\transp{a_{1}}X,...,\transp{a_{j}}X)$ is
$\transp{b_{k}}X=E(\transp{b_{k}}Y)+\transp{b_{k}}\varepsilon$,
where $\varepsilon$ is a centered random vector such that $\transp{b_{k}}\varepsilon$ is orthogonal to $E(\transp{b_{k}}X/\transp{a_{1}}X,...,\transp{a_{j}}X)$.
\end{corollaire}
\section{ Simulations}\label{Simul400}

Let us study four examples. The first involves a $\chi^2$-divergence, the second a Hellinger distance, the third a Cressie-Read divergence (still with $\gamma=1.25$) and the fourth a Kullback Leibler divergence.\\
In each example, our program will follow our algorithm and will aim at creating a sequence of densities $(g^{(j)})$, $j=1,..,k$, $k<d$, such that $g^{(0)}=g,$ $g^{(j)}=g^{(j-1)}f_{a_j}/[g^{(j-1)}]_{a_j}$ and $\Phi(g^{(k)},f)=0,$ with $\Phi$ being a divergence and $a_j=arg\inf_b \Phi(g^{(j-1)}f_b/[g^{(j-1)}]_b,f),$ for all $j=1,...,k$. 
Moreover, in the second example, we will study the robustness of our method with two ouliers. In the third example, defining $(X_1,X_2)$ as a vector with $f$ as density, we will study the regression of $X_1$ on $X_2$. And finally, in the fourth example, we will perform our goodness-of-fit test for copulas.
\begin{Sim}[With the $\chi^2$ divergence]\label{Sim1}
$\\$We are in dimension $3$(=d), and we consider a sample of $50$(=n) values of a random variable $X$ with a density law $f$ defined by :

$f(x)=Gaussian(x1+x2).Gaussian(x0+x2).Gumbel(x0+x1)$,\\
where the Normal law parameters are $(-5,2)$ and $(1,1)$ and where the  Gumbel distribution parameters are $-3$ and $4$. Let us generate then a Gaussian random variable $Y$ - that we will name $g$ - with a density presenting the same mean and variance as $f$.\\
We theoretically obtain $k=1$ and $a_1=(1,1,0)$. To get  this result, we perform the following test:

$H0:\ a_1=(1,1,0)\text{ versus }(H_1):\ a_1\not=(1,1,0).$\\
Then, corollary \ref{KernelLOIDUCRITERE2coro} enables us to estimate $a_1$ by the following 0.9(=$\alpha$) level confidence ellipsoid 

${\mathcal E}_{1}=\{b \in \R^3;\  (Var_{\bf P}(M(b,b)))^{(-1/2)}\Pn_nM(b,b)\leq q^{\cN(0,1)}_{\alpha}/\sqrt n$

\hspace{7.5cm}$\simeq 0,2533/7.0710678=0.03582203\}.$\\
And, we obtain

$\\$\begin{tabular}{cll}
\hline
Our Algorithm &&\\
\hline
\multirow{3}{7cm}{Projection Study 0 :}
 & minimum : 0.0201741    \\  \cline{2-3}
 & at point : (1.00912,1.09453,0.01893) \\  \cline{2-3}
 & P-Value : 0.81131     \\
\hline
\multirow{1}{7cm}{Test :}
 & $H_0$ : $a_1\in {\mathcal E}_{1}$ : True  \\
\hline
\multirow{1}{7cm}{$\chi^2$(Kernel Estimation of $g^{(1)}$, $g^{(1)}$)}
& 6.1726 \\
\hline
\end{tabular}\\
$\\$Therefore, we conclude that $f=g^{(1)}.$
\end{Sim}
\begin{Sim}[With the Hellinger distance $H$]\label{Sim2}
$\\$We are in dimension $20$(=d). We first generate a sample with $100$(=n) observations, namely two outliers $x=(2,0,\ldots,0)$ and $98$ values of a random variable $X$ with a density law $f$ defined by
$f(x)=Gumbel(x_0).Normal(x_1,\ldots,x_9)$, where the Gumbel law parameters are -5 and 1 and where the normal distribution is reduced and centered.\\
Our reasoning is the same as in Simulation \ref{Sim1}. \\In the first part of the  program, we theoretically obtain $k=1$ and $a_1=(1,0,\ldots,0)$. To get  this result, we perform the following test $(H_0):\ a_1=(1,0,\ldots,0)\ versus\ (H_1):\ a_1\not=(1,0,\ldots,0)$. \\
We estimate $a_1$ by the following 0.9(=$\alpha$) level confidence ellipsoid

$\cE_{i}=\{b \in \R^2;\  (Var_\PP(M(b,b)))^{-1/2}\Pn_nM(b,b)\leq q^{\cN(0,1)}_{\alpha}/\sqrt n\simeq 0.02533\}.$\\
And, we obtain\\
\begin{tabular}{cll}
\hline\noalign{\smallskip}
Our Algorithm &&\\
\noalign{\smallskip}\hline\noalign{\smallskip}
\multirow{5}{2cm}{Projection Study 0}
 & minimum : 0.002692 \\  \cline{2-3}
 & at point : (1.01326, 0.0657, 0.0628, 0.1011, 0.0509, 0.1083,\\
 & 0.1261, 0.0573, 0.0377, 0.0794, 0.0906, 0.0356, 0.0012,\\
 & 0.0292, 0.0737, 0.0934,  0.0286, 0.1057, 0.0697, 0.0771)\\ \cline{2-3}
 & P-Value : 0.80554\\
\noalign{\smallskip}\hline
\multirow{1}{2cm}{Test :}
 & $H_0$ : $a_1\in \cE_{1}$ : True \\
\noalign{\smallskip}\hline
\multirow{1}{3.5cm}{$H$(Estimate of $g^{(1)}$, $g^{(1)}$)}
& 3.042174 \\
\noalign{\smallskip}\hline
\end{tabular}
$\\$Therefore, we conclude that $f=g^{(1)}.$
\end{Sim}
\begin{Sim}[With the Cressie-Read divergence ($\Phi$)]\label{Sim3}
$\\$We are in dimension $2$(=d), and we consider a sample of $50$(=n) values of a random variable $X=(X_1,X_2)$ with a density law $f$ defined by  $f(x)=Gumbel(x_0).Normal(x_1)$,
where the Gumbel law parameters are -5 and 1 and where the  normal distribution parameters are $(0,1)$. Let us generate then a Gaussian random variable $Y$ - that we will name $g$ - with a density presenting same mean and variance as $f$.\\
We theoretically obtain $k=1$ and $a_1=(1,0)$. To get  this result, we perform the following test:

$H0:\ a_1=(1,0)\text{ versus }(H_1):\ a_1\not=(1,0).$\\
Then, corollary \ref{KernelLOIDUCRITERE2coro} enables us to estimate $a_1$ by the following 0.9(=$\alpha$) level confidence ellipsoid 

${\mathcal E}_{1}=\{b \in \R^2;\  (Var_{\bf P}(M(b,b)))^{(-1/2)}\Pn_nM(b,b)\leq q^{\cN(0,1)}_{\alpha}/\sqrt n\simeq0.03582203\}.$\\
And, we obtain
$\\$\begin{tabular}{cll}
\hline
Our Algorithm &&\\
\hline
\multirow{3}{9cm}{Projection Study 0 :}
 & minimum : 0.0210058  \\  \cline{2-3}
 & at point : (1.001,0.0014) \\  \cline{2-3}
 & P-Value : 0.989552   \\
\hline
\multirow{1}{9cm}{Test :}
 & $H_0$ : $a_1\in {\mathcal E}_{1}$ : True  \\
\hline
\multirow{1}{9cm}{$\Phi$(Kernel Estimation of $g^{(1)}$, $g^{(1)}$)}
& 6.47617 \\
\hline
\end{tabular}\\
$\\$Therefore, we conclude that $f=g^{(1)}$.
\end{Sim}
\begin{figure}[htb!]
 \center
 \includegraphics[scale=0.52, angle=-90]{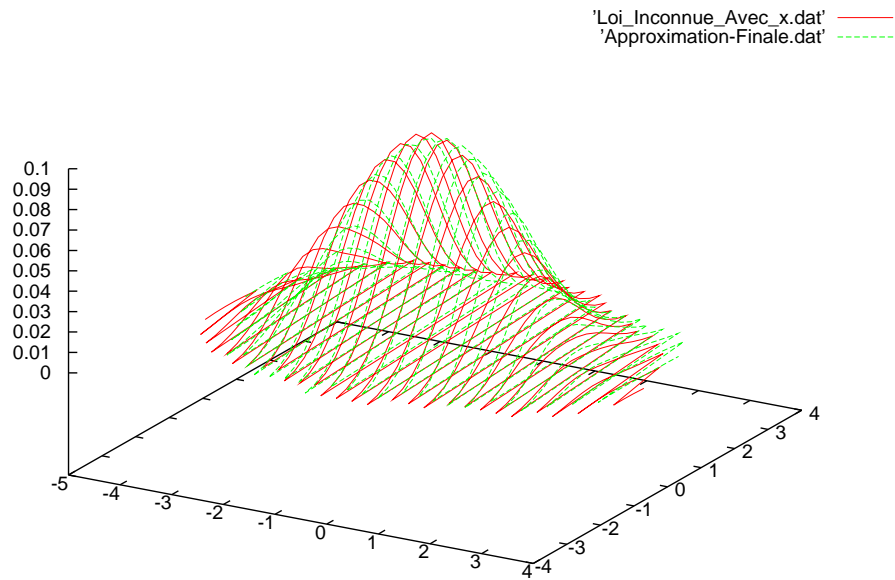}
 \caption{\it Graph of the distribution to estimate (red) and of our own estimate (green).}
\end{figure}
\begin{figure}[htb!]
 \center
 \includegraphics[scale=0.52, angle=-90]{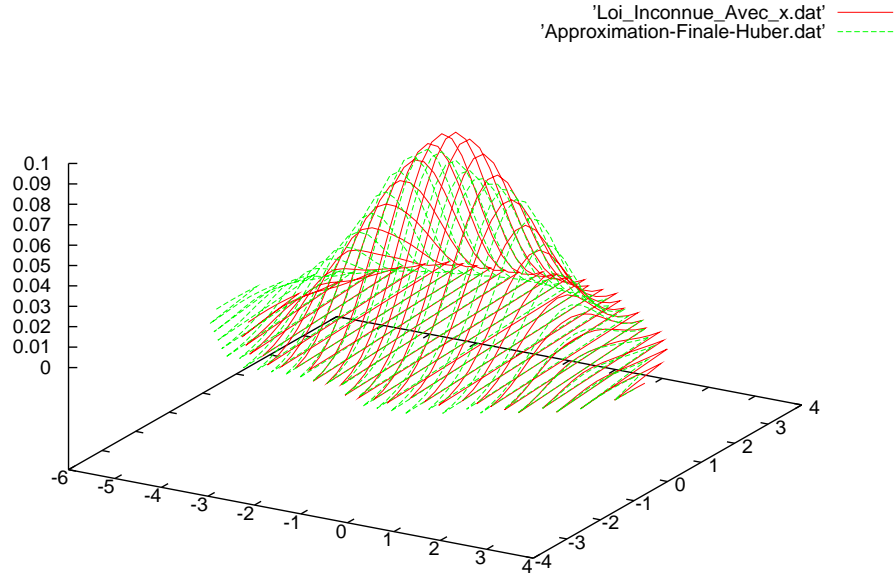}
 \caption{\it Graph of the distribution to estimate (red) and of Huber's estimate (green).}
\end{figure}
\newpage
At present, keeping the notations of this simulation, let us study the regression of $X_1$ on $X_2$.\\
Our algorithm leads us to infer that the density of $X_1$ given $X_2$ is the same as the density of $y_1$ given $Y_2$. Moreover, property \ref{Phimini} implies that the co-factors of $f$ are the same with all divergence. Consequently, we can use  theorem \ref{FirstReg}, i.e. it implies that $X_1=E(Y_1/Y_2)+\varepsilon,$ where $\varepsilon$ is a centered random variable orthogonal to $E(X_1/X_2)$.
Thus, since $g$ is a Gaussian density, remark \ref{FirstRegRem} implies that
$$
X_1=E(Y_1)+\frac{Cov(Y_1,Y_2)}{Var(Y_2)}(Y_2-E(Y_2))+\varepsilon.
$$
Now, using the least squares method, we estimate $\alpha_1$ and $\alpha_2$ such that 
$X_1=a_1+a_2X_2+\varepsilon.$\\
Thus, the following table presents the results of our regression and of the least squares method if we assume that $\varepsilon$ is Gaussian.\\

\noindent \begin{tabular}{clll}
\hline
\multirow{6}{5cm}{Our Regression}
 &$E(Y_1)$      &   -4.545483  \\  \cline{2-3}
 &$Cov(Y_1,Y_2)$&   0.0380534  \\  \cline{2-3}
 &$Var(Y_2)$    &   0.9190052  \\  \cline{2-3}
 &$E(Y_2)$      &   0.3103752  \\  \cline{2-3}
 &correlation coefficient $(Y_1,Y_2)$ &  0.02158213   \\
\hline
\multirow{3}{5cm}{Least squares method}
 &$a_1$                               &  -4.34159227   \\  \cline{2-3}
 &$a_2$                               &  0.06803317   \\  \cline{2-3}
 &correlation coefficient $(X_1,X_2)$ &  0.04888484   \\
\hline
\end{tabular}\\
\begin{figure}[htb!]\label{RegImage}
 \center
 \includegraphics[scale=0.52]{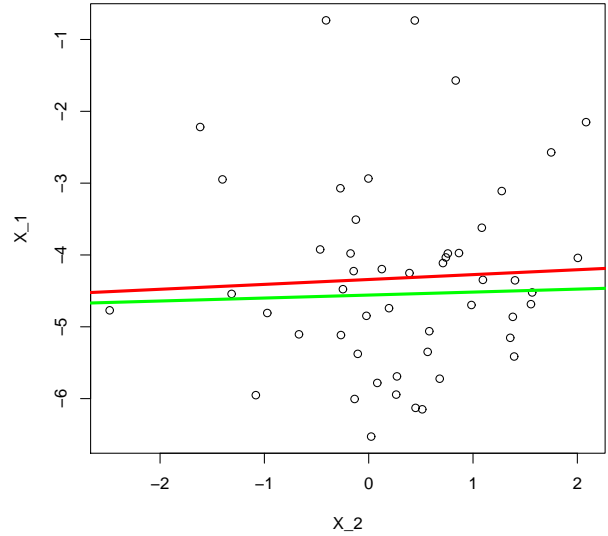}
 \caption{\it Graph of the regression of X1 on X2 based on the least squares method (red) and based on our theory (green).}
\end{figure}
\begin{Sim}[With the relative entropy $K$]\label{Sim55555}
$\\$We are in dimension $2$(=d), and we use the relative entropy to perform our optimisations. Let us consider a sample of $50$(=n) values of a random variable $X$ with a density law $f$ defined by :

$f(x)=c_\rho(F_{Gumbel}(x_0),F_{Exponential}(x_1)).Gumbel(x_0).Exponential(x_1)$,\\
where :\\
$\bullet$ $c$ is the Gaussian copula with correlation coefficient $\rho=0.5$,\\
$\bullet$ the  Gumbel distribution parameters are $-1$ and $1$ and\\
$\bullet$  the Exponential density parameter is $2$. \\
Let us generate then a Gaussian random variable $Y$ - that we will name $g$ - with a density presenting the same mean and variance as $f$.\\
We theoretically obtain $k=2$ and $(a_1,a_2)=((1,0),(0,1))$. To get  this result, we perform the following test:

$(H_0):\ (a_1,a_2)=((1,0),(0,1))\text{ versus }(H_1):\ (a_1,a_2)\not=((1,0),(0,1)).$\\
Then, theorem \ref{TestCop} enables us to verify $(H_0)$ by the following 0.9(=$\alpha$) level confidence ellipsoid \\
${\mathcal E}_{2}=\{b \in \R^2;(Var_{\bf P}(M(b,b)))^{(-1/2)}\Pn_nM(b,b)\leq q^{\cN(0,1)}_{\alpha}/\sqrt n\simeq 0,2533/7.0710678 =0.0358220\}.$\\
And, we obtain\\
\begin{tabular}{cll}
\hline
Our Algorithm &&\\
\hline
\multirow{3}{8cm}{Projection Study number 0 :}
 & minimum :  0.445199  \\  \cline{2-3}
 & at point : (1.0142,0.0026) \\  \cline{2-3}
 & P-Value :  0.94579   \\
\hline
\multirow{1}{8cm}{Test :}
 & $H_0$ : $a_1\in {\mathcal E}_{1}$ : False  \\
\hline
\end{tabular}\\
\begin{tabular}{cll}
&&\\
\hline
\multirow{3}{8cm}{Projection Study number 1 :}
 & minimum :  0.0263    \\  \cline{2-3}
 & at point : (0.0084,0.9006) \\  \cline{2-3}
 & P-Value :  0.97101    \\
\hline
\multirow{1}{8cm}{Test :}
 & $H_0$ : $a_2\in {\mathcal E}_{2}$ : True  \\
\hline
\multirow{1}{8cm}{$K$(Kernel Estimation of $g^{(2)}$, $g^{(2)}$)}
& 4.0680 \\
\hline
\end{tabular}

Therefore, we can conclude that $H_0$ is verified.
\begin{figure}[htb!]
\center
\includegraphics[scale=0.52]{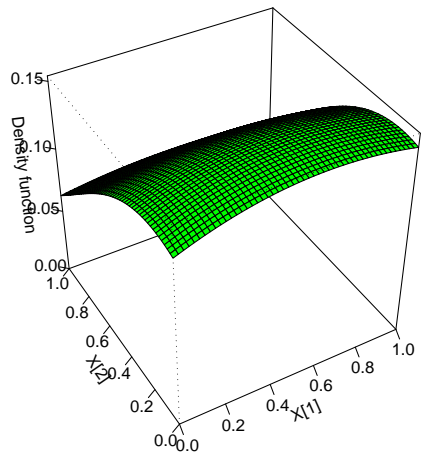}
\caption{\it Graph of the estimate of $(x_0,x_1)\mapsto c_\rho(F_{Gumbel}(x_0),F_{Exponential}(x_1))$.}
\end{figure}
\end{Sim}
\noindent {\bf Critics of the simulations}

In the case where $f$ is unknown, we will never be sure to have reached the minimum of the $\Phi$-divergence:  we have indeed used the simulated annealing method to solve our optimisation problem, and therefore it is only when the number of random jumps tends in theory towards infinity that the probability to reach the minimum tends to 1.
We also note that no theory on the optimal number of jumps to implement does exist, as this number depends on the specificities of each particular problem.\\
Moreover, we choose the $50^{-\frac{4}{4+d}}$ (resp. $100^{-\frac{4}{4+d}}$) for the AMISE of simulations \ref{Sim1}, \ref{Sim2} and \ref{Sim3} (resp. simulation \ref{Sim55555}). This choice leads us to simulate 50 (resp. 100) random variables - see \cite{MR1191168} page 151 -, none of which have been discarded to obtain the truncated sample.\\
Finally, we remark that some of the key advantages of our method over Huber's consist in the fact that - since there exist divergences smaller than the relative entropy - our method requires a considerably shorter computation time and also in the in the superiority in robustness of our method.
\newpage 
\noindent {\bf  Conclusion}

Projection Pursuit is useful in evidencing characteristic structures as well as one-dimensional projections and their associated distributions in multivariate data.
\cite{MR790553} shows us how to achieve it through maximization of the relative entropy.\\ 
The present article shows that our $\Phi$-divergence method constitutes a good alternative to Huber's particularly in terms of regression and robustness as well as in terms of copula's study. Indeed, the convergence results and simulations we carried out, convincingly fulfilled our expectations regarding our methodology.
\appendix
\section{Reminders}
\subsection{ $\Phi$-Divergence}\label{FiDiv}
Let us call $h_a$ the density of  $\transp a Z$ if  $h$ is the density of $Z$. 
Let $\varphi$ be a strictly convex function defined by $\varphi: \overline{\R^+}\tend\overline{\R^+},$ and such that $\varphi(1)=0$.
\begin{definition}
We define the $\Phi-$divergence of $P$ from $Q$, where $P$ and $Q$ are two probability distributions over a space $\Omega$ such that $Q$ is absolutely continuous with respect to $P$, by
\begin{equation}\label{def:div}
\Phi(Q,P)=\int\varphi(\frac{dQ}{dP})dP.
\end{equation}
The above expression (\ref{def:div}) is also valid if $P$ and $Q$ are both dominated by the same probability.
\end{definition}
The most used distances (Kullback, Hellinger or $ \chi^2$) belong to the Cressie-Read family \\(see \cite{CressieRead}, \cite{Csiszar67} and the books of \cite{MR926905}, \cite{MR2183173} and \cite{MR1075502}). They are defined by a specific $\varphi$. Indeed,\\
- with the relative entropy, we associate $\varphi(x)=xln(x)-x+1$\\
- with the Hellinger distance, we associate $\varphi(x)=2(\sqrt x-1)^2$\\ 
- with the $\chi^2$ distance, we associate $\varphi(x)=\frac{1}{2}(x-1)^2$\\
- more generally, with power divergences, we associate $\varphi(x)=\frac{x^\gamma-\gamma x+\gamma-1}{\gamma(\gamma-1)}$, where $\gamma\in\R\setminus {(0,1)}$\\
- and, finally, with the $L^1$ norm, which is also a divergence, we associate $\varphi(x)=|x-1|.$\\
In particular we have the following inequalities: 

$\ \ \ \ \ \ \ \ \ $ $d_{L^1}(g,f)\leq K(g,f)\leq \chi^2(g,f)$.
$\\$Let us now present some well-known properties of divergences.
\begin{PROPIE}\label{Phimini}
We have $\Phi(P,Q)=0\Leftrightarrow P=Q.$
\end{PROPIE}
\begin{PROPIE}\label{K-SCI}
The application $Q\mapsto \Phi(Q,P)$ is greater than the $L^1$ distance, convex, lower semi-continuous (l.s.c.) - for the topology that makes all the applications of the form $Q\mapsto\int fdQ$ continuous where $f$ is bounded and continuous - as well as l.s.c. for the topology of the uniform convergence.
\end{PROPIE}
\begin{PROPIE}[corollary  (1.29), page 19 of \cite{MR926905}]\label{ExitenceDeLEntropieDesProj}
If $T:(X,A)\to (Y,B)$ is measurable and if $K(P,Q)<\infty,$ then $K(P,Q)\geq K(PT^{-1},QT^{-1}),$
with equality being reached when $T$ is surjective for $(P,Q)$.
\end{PROPIE}
\begin{theoreme}[theorem III.4 of \cite{AZE97}]\label{azeIII4}
Let $f:I\to \R$ be a convex function.  Then $f$ is a Lipschitz function in all compact intervals $[a,b]\subset int\{I\}.$ In particular, $f$ is continuous on $int\{I\}$.
\end{theoreme}
\subsection{ Useful lemmas}

Through a reductio ad absurdum argument, we derive lemmas \ref{ProjBorne} and \ref{LoiCondBorne} :
\begin{lemme} \label{ProjBorne}
Let $f$ be a density in $\R^d$ bounded and positive.  Then, any projection density of $f$ - that we will name $f_a$, with $a\in\R^d_*$ - is also bounded and positive in $\R$.
\end{lemme}
\begin{lemme} \label{LoiCondBorne}
Let $f$ be a density in $\R^d$ bounded and positive.  Then any density $f(./\transp ax)$, for any $a\in\R^d_*$, is also bounded and positive.
\end{lemme}
\noindent By induction and from lemmas \ref{ProjBorne} and \ref{LoiCondBorne}, we have
\begin{lemme}\label{GkBorne}
If $f$ and $g$ are positive and bounded densities, then $g^{(k)}$ is positive and bounded.
\end{lemme}
\noindent Finally we introduce a last lemma
\begin{lemme}\label{aleph}
Let $f$ be an absolutely continuous density, then, for all sequences $(a_n)$ tending  to $a$ in $\R^d_*$, sequence $f_{a_n}$ uniformly converges towards $f_a$.
\end{lemme}
\begin{proof}
For all $a$ in $\R^d_*$, let $F_a$ be the cumulative distribution function of $\transp aX$ and $\psi_a$ be a complex function defined by $\psi_a(u,v)=F_a(\cR e(u+iv))+iF_a(\cR e(v+iu))$, for all $u$ and $v$ in $\R$.\\First, the function $\psi_a(u,v)$ is an analytic function, because $x\mapsto f_a(\transp ax)$ is continuous and as a result of the corollary of Dini's second theorem - according to which 
{\it "A sequence of cumulative distribution functions which pointwise converges on $\R$ towards a continuous cumulative distribution function $F$ on $\R$, uniformly converges towards $F$ on $\R$"}-
we deduct that, for all sequences $(a_n)$ converging towards $a$, $\psi_{a_n}$ uniformly converges towards $\psi_a$.
Finally, the Weierstrass theorem, (see proposal $(10.1)$ page 220 of the "Calcul infinitésimal" book of Jean Dieudonné), implies that all sequences $\psi_{a,n}'$ uniformly converge towards $\psi_a'$, for all $a_n$ tending to $a$. We can therefore conclude.
\end{proof}
\section{ Study of the sample}\label{truncSample}
Let $X_1$, $X_2$,..,$X_m$ be a sequence of independent random vectors with same density  $f$.
Let $Y_1$, $Y_2$,..,$Y_m$ be a sequence of independent random vectors with same density  $g$.
Then, the kernel estimators $f_m$, $g_m$, $f_{a,m}$ and $g_{a,m}$ of $f$, $g$, $f_{a}$ and $g_{a}$, for all $a\in\R^d_*$, almost surely and uniformly converge since we assume that the bandwidth $h_m$ of these estimators meets the following conditions (see \cite{MR0345296}): 

$(\mathcal Hyp)$: $h_m\searrow_m0$, $mh_m\nearrow_m\infty$, $mh_m/L(h_m^{-1})\to_m\infty$ and $L(h_m^{-1})/LLm\to_m\infty$,\\ with $L(u)=ln(u\vee e)$.\\
Let us consider 

$B_1(n,a)=\frac{1}{n}\Sigma_{i=1}^n\varphi'\{\frac{f_{a,n}(\transp aY_i)}{g_{a,n}(\transp aY_i) }\frac{g_n(Y_i)}{f_n(Y_i)}\}\frac{f_{a,n}(\transp aY_i)}{g_{a,n}(\transp aY_i) }$ and
$B_2(n,a)=\frac{1}{n}\Sigma_{i=1}^n\varphi^*\{\varphi'\{\frac{f_{a,n}(\transp aX_i)}{g_{a,n}(\transp aX_i) }\frac{g_n(X_i)}{f_n(X_i}\}\}.$\\
Our goal is to estimate the minimum of $\Phi(g\frac{f_a}{g_a},f)$.
To do this, it is necessary for us to truncate our samples:\\
Let us consider now a positive sequence $\theta_m$ such that $\theta_m\to 0,\ y_m/\theta_n^2\to 0,$ where $y_m$ is the almost sure convergence rate of the kernel density estimator - $y_m=O_\PP(m^{-\frac{2}{4+d}})$, see lemma \ref{KernRate} - $y^{(1)}_m/\theta_m^2\to 0,$ where $y^{(1)}_m$ is defined by $$|\varphi(\frac{g_m(x)}{f_m(x)}\frac{f_{b,m}(\transp bx)}{g_{b,m}(\transp bx)})-\varphi(\frac{g(x)}{f(x)}\frac{f_b(\transp bx)}{g_b(\transp bx)})|\leq y^{(1)}_m$$ for all $b$ in $\R^d_*$ and all $x$ in $\R^d$, and finally  $\frac{y^{(2)}_m}{\theta_m^2}\to 0,$ where $y^{(2)}_n$ is defined by $$|\varphi'(\frac{g_m(x)}{f_m(x)}\frac{f_{b,m}(\transp bx)}{g_{b,m}(\transp bx)})-\varphi'(\frac{g(x)}{f(x)}\frac{f_b(\transp bx)}{g_b(\transp bx)})|\leq y^{(2)}_m$$ for all $b$ in $\R^d_*$ and all $x$ in $\R^d$.\\
We will generate $f_m$, $g_m$ and $g_{b,m}$ from the starting sample and we will select the $X_i$ and $Y_i$ vectors such that $f_m(X_i)\geq \theta_m$ and $g_{b,m}(\transp bY_i)\geq \theta_m$, for all $i$ and for all $b\in \R^d_*$. \\
The vectors meeting these conditions will be called $X_1,X_2,...,X_n$ and $Y_1,Y_2,...,Y_n$.\\
Consequently, the next proposition provides us with the condition required for us to derive our estimations
\begin{proposition}\label{QuotientDonneLoi}
Using the notations introduced in \cite{MR2054155} and in section \ref{HypoF},
it holds $\lim_{n\to\infty}\sup_{a\in\R^d_*}|(B_1(n,a)-B_2(n,a))-\Phi(g\frac{f_a}{g_a},f)|=0.$
\end{proposition}
\begin{remarque}\label{Scott}
With the relative entropy, we can  take for $\theta_m$ the expression $m^{-\nu}$, with $0<\nu<\frac{1}{4+d}$.
\end{remarque}
\section{ Case study : $f$ is known}\label{fKnown}
In this Annex, we will study the case when $f$ and $g$ are known. We will then use the notations introduced in sections \ref{HypoF} and \ref{Estimofa1} with $f$ and $g$, i.e. no longer with their kernel estimates.
\subsection{ Convergence study and Asymptotic Inference at the  $k^{\text{th}}$ step of the algorithm}
In this paragraph, when $k$ is less than or equal to $d$, we will show that the sequence $(\check a_k)_n$ converges towards $a_k$ and that the sequence $(\check g^{(k)})_n$ converges towards $g^{(k)}$. \\
Both $\check \gamma_n$ and $\check c_n(a)$ are M-estimators and estimate $a_k$ - see \cite{MR2054155}.
We state
\begin{proposition}\label{pConv2}
Assuming $(H1)$ to $(H3)$ hold. Both $\sup_{a\in\Theta}\|\check c_n(a)-a_k\|$ and $\check \gamma_n $ tends to $a_k$ a.s.
\end{proposition}
\noindent Finally, the following theorem shows us that $\check  g^{(k)}$ converges uniformly almost everywhere towards $g^{(k)}$, for any $k=1..d$.
\begin{theoreme}\label{KRessultatPricipal}
Assumimg $(H1)$ to $(H3)$ hold. Then, $\check  g^{(k)}\to_n   g^{(k)}$ a.s. and uniformly a.e.
\end{theoreme}
The following theorem shows that $\check g^{(k)}$ converges at the rate $O_{\PP}(n^{-1/2})$ in three differents cases, namely for any given $x$, with the  $L^1$ distance and with the $\Phi-$divergence:
\begin{theoreme}\label{Superdiffconv}Assuming  $(H0)$ to $(H3)$ hold, for any $k=1,...,d$ and any $x\in\R^d$, we have
\begin{eqnarray}
&&|\check g^{(k)}(x)-g^{(k)}(x)|= O_{\PP}(n^{-1/2}),\label{diffconv1}\\
&&\int |\check g^{(k)}(x)-g^{(k)}(x)|dx=O_{\PP}(n^{-1/2}),\label{diffconv2}\\
&&|K(\check g^{(k)},f)-K(g^{(k)},f)|=O_{\PP}(n^{-1/2}).\label{diffconv3}
\end{eqnarray}
\end{theoreme}
The following theorem shows that the laws of our estimators of $a_k$, namely $\check c_n(a_k)$ and $\check \gamma_n$, converge towards a linear combination of Gaussian variables.
\begin{theoreme}\label{filoiestimateurs}
Assuming that conditions $(H1)$ to $(H6)$ hold, then\\
$\sqrt n\cA.(\check c_n(a_k)-a_k)\cvL \cB.\cN_d(0,\PP\|\frac{\dr}{\dr b}M(a_k,a_k)\|^2)+\cC.\cN_d(0,\PP\|\frac{\dr}{\dr a}M(a_k,a_k)\|^2)$ and\\
$\sqrt n\cA.(\check \gamma_n-a_k)\cvL \cC.\cN_d(0,\PP\|\frac{\dr}{\dr b}M(a_k,a_k)\|^2)+\cC.\cN_d(0,\PP\|\frac{\dr}{\dr a}M(a_k,a_k)\|^2)$\\
where $\cA=(\PP\frac{\dr^2}{\dr b\dr b}M(a_k,a_k)(\PP\frac{\dr^2}{\dr a_i\dr a_j}M(a_k,a_k)+\PP\frac{\dr^2}{\dr a_i\dr b_j}M(a_k,a_k)))$, \\
$\cC=\PP\frac{\dr^2}{\dr b\dr b}M(a_k,a_k)$ and
$\cB=\PP\frac{\dr^2}{\dr b\dr b}M(a_k,a_k)+\PP\frac{\dr^2}{\dr a_i\dr a_j}M(a_k,a_k)+\PP\frac{\dr^2}{\dr a_i\dr b_j}M(a_k,a_k).$
\end{theoreme}

\subsection{A stopping rule for the procedure}

We now assume that the algorithm does not stop after $d$ iterations. We then remark that, it still holds - for any $i>d$: 

$\bullet$  $g^{(i)}(x)=g(x)\Pi_{k=1}^i\frac{f_{ a_k}(\transp { a_k}x)}{[g_n^{(k-1)}]_{ a_k}(\transp { a_k}x)}$, with $g^{(0)}=g$.

$\bullet$  $K(g^{(0)},f)\geq K(g^{(1)},f)\geq K(g^{(2)},f)...\geq 0$.

$\bullet$ Theorems \ref{KRessultatPricipal}, \ref{Superdiffconv} and \ref{filoiestimateurs}.\\
Moreover, as explained in section 14 of \cite{MR790553} for the relative entropy, the sequence $( \Phi(g^{(k-1)}\frac{f_{a_k}}{g^{(k-1)}_{a_k}},f))_{k\geq 1}$ converges towards zero. Then, in this paragraph, we will show that $g^{(i)}$ converges towards $f$ in $i$. And finally, we will provide a stopping rule for this identification procedure.
\subsubsection{ Representation of $f$}
Under $(H0)$, the following proposition shows us that the probability measure with density $g^{(k)}$ converges towards the probability measure with density $f$ : 
\begin{proposition}[Representation of $f$]\label{cvl}
We have  $\lim_k g^{(k)}= f$ a.s.
\end{proposition}
\subsubsection{ Testing of the criteria}
Through a test of the criteria, namely $a\mapsto \Phi(g^{(k-1)}\frac{f_a}{g^{(k-1)}_a},f)$, we will build a stopping rule for this procedure. First, the next theorem enables us to derive the law of the  criteria.
\begin{theoreme} \label{LOIDUCRITERE} 
Assuming that $(H1)$ to $(H3)$, $(H6)$ and $(H8)$ hold. Then,

$\sqrt n(Var_{\PP}(M(\check c_n(\check \gamma_n),\check \gamma_n)))^{-1/2}(\Pn_nM(\check c_n(\check \gamma_n),\check \gamma_n)-\Pn_nM(a_k,a_k)) \cvL \cN(0,I)$,\\
where $k$ represents the $k^{th}$ step of the  algorithm and with $I$ being the identity matrix in $\R^d$.
\end{theoreme}
\noindent Note that $k$ is fixed in theorem \ref{LOIDUCRITERE} since $\check \gamma_n   =\ arg\inf_{a\in\Theta }\ \sup_{c\in\Theta }\ \Pn_nM(c,a)$ where $M$ is a known function of $k$ - see section \ref{HypoF}. Thus, in the case where $\Phi(g^{(k-1)}\frac{f_{a_k}}{g^{(k-1)}_{a_k}},f)= 0$, we obtain
\begin{corollaire} \label{LOIDUCRITERE2} 
Assuming that $(H1)$ to $(H3)$, $(H6)$, $(H7)$ and $(H8)$ hold. Then, 

$\sqrt n(Var_{\PP}(M(\check c_n(\check \gamma_n),\check \gamma_n)))^{-1/2}(\Pn_nM(\check c_n(\check \gamma_n),\check \gamma_n)) \cvL \cN(0,I).$
\end{corollaire}
\noindent Hence, we propose the test of the null hypothesis 

$(H_0)$ : $K(g^{(k-1)}\frac{f_{a_k}}{g^{(k-1)}_{a_k}},f)= 0$ versus $(H_1)$ : $K(g^{(k-1)}\frac{f_{a_k}}{g^{(k-1)}_{a_k}},f)\not= 0$.\\
Based on this result, we stop the algorithm, then, defining $a_k$ as the last vector generated, we derive from corollary \ref{LOIDUCRITERE2} a $\alpha$-level confidence ellipsoid around $a_k$, namely

$\cE_k=\{b\in\R^d;\ \sqrt n(Var_{\PP}(M(b,b)))^{-1/2}\Pn_nM(b,b)\leq q_{\alpha}^{\cN(0,1)} \}$,\\
where $q_{\alpha}^{\cN(0,1)}$ is the quantile of a $\alpha$-level reduced centered normal distribution.\\ Consequently, the following corollary provides us with a confidence region for the above test:
\begin{corollaire}\label{LOIDUCRITERE2coro}
$\cE_k$ is a confidence region for the test of the null hypothesis $(H_0)$ versus $(H_1)$.
\end{corollaire}
\section{ The first co-vector of $f$ simultaneously optimizes four problems}\label{a1solve4}
Let us first study Huber's analytic approach.\\
Let $\cR'$ be the class of all positive functions $r$ defined on $\R$ and such that $f(x)r^{-1}(\transp ax)$ is a density on $\R^d$  for all $a$ belonging to $\R^d_*$. The following proposition shows that there exists a vector $a$ such that  $\frac{f_a}{g_a}$ minimizes $K(fr^{-1},g)$ in $r$:
\begin{proposition}[Analytic Approach]\label{lemmeHuber0prop}
There exists a vector $a$ belonging to $\R^d_*$ such that 

$arg\min_{r\in\cR'}K(fr^{-1},g)=\frac{f_a}{g_a},\text{ and } r(\transp ax)=\frac{f_a(\transp ax)}{g_a(\transp ax)}$ as well as $  K(f,g)=  K(f_a,g_a)+  K(f\frac{g_a}{f_a},g).$
\end{proposition}
\noindent Let us also study Huber's synthetic approach:\\
Let $\cR$ be the class of all positive functions $r$ defined on $\R$ and such that $g(x)r(\transp ax)$ is a density on $\R^d$ for all $a$ belonging to $\R^d_*$. The following proposition shows that there exists a vector $a$ such that  $\frac{f_a}{g_a}$ minimizes $K(gr,f)$ in $r$:
\begin{proposition}[Synthetic Approach]\label{lemmeHuberprop}
There exists a vector $a$ belonging to $\R^d_*$ such that

$arg\min_{r\in\cR}K(f,gr)=\frac{f_a}{g_a},\text{ and } r(\transp ax)=\frac{f_a(\transp ax)}{g_a(\transp ax)}$ as well as $  K(f,g)=  K(f_a,g_a)+  K(f,g\frac{f_a}{g_a}).$
\end{proposition}
\noindent In the meanwhile, the following proposition shows  that there exists a vector $a$ such that   $\frac{f_a}{g_a}$ minimizes  $K(g,fr^{-1})$ in $r$.
\begin{proposition} \label{liendeminimaxi}
There exists a vector $a$ belonging to $\R^d_*$ such that

$arg\min_{r\in\cR'}K(g,fr^{-1})=\frac{f_a}{g_a},\text{ and } r(\transp ax)=\frac{f_a(\transp ax)}{g_a(\transp ax)}$ as well as  $K(g,f)=  K(g_a,f_a)+K(g,f\frac{g_a}{f_a})$.
\end{proposition}
\begin{remarque}\label{criteria-H}First, through property \ref{ExitenceDeLEntropieDesProj}  page \pageref{ExitenceDeLEntropieDesProj}, we get $K(f,g\frac{f_a}{g_a})= K(g,f\frac{g_a}{f_a})= K(f\frac{g_a}{f_a},g)$ and $K(f_a,g_a)=K(g_a,f_a)$. Thus, proposition \ref{liendeminimaxi} implies that finding the argument of the maximum of $K(g_a,f_a)$ amounts to finding the argument of the maximum $K(f_a,g_a)$. Consequently, the criteria of Huber's methodologies is $a\mapsto K(g_a,f_a)$.
Second, if the $\Phi$-divergence is the relative entropy, then our criteria is $a\mapsto K(g\frac{g_a}{f_a},f)$ and property \ref{ExitenceDeLEntropieDesProj} implies $K(g,f\frac{g_a}{f_a})= K(g\frac{f_a}{g_a},f)$.
\end{remarque}
To recapitulate, the choice of $r=\frac{f_a}{g_a}$ enables us to simultaneously solve the following four optimisation problems, for $a\in\R^d_*$:

First, find $a$ such that $a\ =\ {\text{\itshape arginf}}_{a\in\R^d_*}\ K(f\frac{g_a}{f_a},g),$

Second, find $a$ such that $a\ =\ {\text{\itshape arginf}}_{a\in\R^d_*}\ K(f,g\frac{f_a}{g_a}),$

Third, find $a$ such that $a\ =\ {\text{\itshape argsup}}_{a\in\R^d_*}\ K(g_a,f_a),$

Fourth, find $a$ such that $a\ =\ {\text{\itshape arginf}}_{a\in\R^d_*}\ K(g\frac{f_a}{g_a},f).$
\section{Hypotheses' discussion}\label{DiscussHyp}
\subsection{ Discussion of $(H2)$.}
\noindent Let us work with the relative entropy and with $g$ and $a_1$. \\
For all $b\in\R^d_*$, we have
$\int\varphi^*(\varphi'(\frac{g(x)f_b(\transp bx)}{f(x)g_b(\transp bx)})) f(x)dx=\int (\frac{g(x)f_b(\transp bx)}{f(x)g_b(\transp bx)}-1)f(x)dx=0,$
since, for any $b$ in $\R^d_*$, the function $x\mapsto g(x)\frac{f_b(\transp bx)}{g_b(\transp bx)}$ is a density.
The complement of $\Theta^\Phi$ in $\R^d_*$ is $\emptyset$ and then the supremum looked for in $\overline{\R}$ is $-\infty$.
We can therefore conclude.
It is interesting to note that we obtain the same verification with $f$, $g^{(k-1)}$ and $a_k$.
\subsection{ Dicussion of $(H4)$.}
\noindent{\it This hypothesis consists in the following assumptions:\\
$\bullet$ We work with the relative entropy, (0)\\
$\bullet$ We have $f(./\transp {a_1}x)=g(./\transp {a_1}x)$, i.e. $K(g\frac{f_1}{g_1},f)=0$ - we could also derive the same proof with $f$, $g^{(k-1)}$ and $a_k$ - (1)}\\
\noindent{\it Preliminary $(A)$:
Shows that
$A=\{(c,x)\in\R^d_*\backslash \{a_1\}\times R^d;\ \frac{f_{a_1}(\transp {a_1}x)}{g_{a_1}(\transp {a_1}x)}>\frac{f_{c}(\transp cx)}{g_{c}(\transp cx)},\ g(x)\frac{f_{c}(\transp cx)}{g_{c}(\transp cx)}> f(x)\}=\emptyset$
through a reductio ad absurdum, i.e. if we assume $A\not=\emptyset$.}\\
Thus, our hypothesis enables us to derive

$f(x)=f(./\transp {a_1}x)f_{a_1}(\transp {a_1}x)=g(./\transp {a_1}x)f_{a_1}(\transp {a_1}x)> g(./\transp {c}x)f_{c}(\transp {c}x)> f$\\
since $\frac{f_{a_1}(\transp {a_1}x)}{g_{a_1}(\transp {a_1}x)}\geq\frac{f_{c}(\transp cx)}{g_{c}(\transp cx)}$ implies $g(./\transp {a_1}x)f_{a_1}(\transp {a_1}x)=g(x)\frac{f_{a_1}(\transp {a_1}x)}{g_{a_1}(\transp {a_1}x)}\geq g(x)\frac{f_{c}(\transp cx)}{g_{c}(\transp cx)}=g(./\transp {c}x)f_{c}(\transp {c}x)$, i.e. $f>f$. We can therefore conclude.
$\\${\it Preliminary $(B)$:
Shows that $B=\{(c,x)\in\R^d_*\backslash \{a_1\}\times R^d;\ \frac{f_{a_1}(\transp {a_1}x)}{g_{a_1}(\transp {a_1}x)}<\frac{f_{c}(\transp cx)}{g_{c}(\transp cx)},\ g(x)\frac{f_{c}(\transp cx)}{g_{c}(\transp cx)}< f(x)\}=\emptyset$
through a reductio ad absurdum, i.e. if we assume $B\not=\emptyset$.}\\
Thus, our hypothesis enables us to derive

$f(x)=f(./\transp {a_1}x)f_{a_1}(\transp {a_1}x)=g(./\transp {a_1}x)f_{a_1}(\transp {a_1}x)< g(./\transp {c}x)f_{c}(\transp {c}x)< f$\\
We can therefore conclude as above.
$\\$Let us now verify  $(H4)$:\\
We have $P M(c,a_1)- P M(c,a)=\int ln(\frac{g(x)f_{c}(\transp cx)}{g_{c}(\transp cx)f(x)})\{\frac{f_{a_1}(\transp {a_1}x)}{g_{a_1}(\transp {a_1}x)}-\frac{f_{c}(\transp cx)}{g_{c}(\transp cx)}\}g(x)dx.$
Moreover, the logarithm $ln$ is negative on 
$\{x\in\R^d_*;\ \frac{g(x)f_{c}(\transp cx)}{g_{c}(\transp cx)f(x)}<1\}$ and is positive on $\{x\in\R^d_*;\ \frac{g(x)f_{c}(\transp cx)}{g_{c}(\transp cx)f(x)}\geq1\}$.\\
Thus, the preliminary studies $(A)$ and $(B)$ show that $ln(\frac{g(x)f_{c}(\transp cx)}{g_{c}(\transp cx)f(x)})$ and $\{\frac{f_{a_1}(\transp {a_1}x)}{g_{a_1}(\transp {a_1}x)}-\frac{f_{c}(\transp cx)}{g_{c}(\transp cx)}\}$  always present a negative product. We can therefore conclude, since $(c,a)\mapsto P M(c,a_1)- P M(c,a)$ is not null for all $c$ and for all $a$ - with $a\not=a_1$.
\section{Proofs}
This last section includes the proofs of most of the lemmas, propositions, theorems and corollaries contained in the present article.
\begin{remarque}\label{GkBor}
1/ $(H0)$ - according to which $f$ and $g$ are assumed to be positive and bounded - through lemma \ref{GkBorne} (see page \pageref{GkBorne}) implies that $\check g^{(k)}$ and $\hat g^{(k)}$ are positive and bounded.\\
2/ remark \ref{implyEstimBounded} page \pageref{implyEstimBounded} implies that $f_n$, $g_n$, $\check g^{(k)}$ and $\hat g^{(k)}$ are positive and bounded since we consider a Gaussian kernel.
\end{remarque}
\noindent{\bf Proof of propositions \ref{lemmeHuber0prop} and \ref{lemmeHuberprop}.}
Let us first study proposition \ref{lemmeHuberprop}.\\
Without loss of generality, we will prove this proposition with $x_1$ in lieu of $\transp aX$.\\
Let us define $g^*=gr$. We remark that $g$ and $g^*$ present the same density conditionally to $x_1$.
Indeed, $g^*_1(x_1)= \int g^*(x)dx_2...dx_d= \int r(x_1)g(x)dx_2...dx_d= r(x_1)\int g(x)dx_2...dx_d= r(x_1)g_1(x_1)$.\\
Thus, we can demonstrate this proposition.\\
We have $g(.|x_1)=\frac{g(x_1,..., x_n)}{g_1(x_1)}$ and $g_1(x_1)r(x_1)$ is the marginal density of $g^*.$
Hence,\\ $\int g^*dx=\int g_1(x_1)r(x_1)g(.|x_1)dx=\int g_1(x_1)\frac{f_1(x_1)}{g_1(x_1)}(\int g(.|x_1)dx_2..dx_d)dx_1=\int f_1(x_1)dx_1=1$ and since $g^*$ is positive, then $g^*$ is a density.
Moreover,
\begin{eqnarray}
  K(f,g^*)&=& \int f\{ln(f)-ln(g^*)\}dx,\label{dernière-1}\\
        &=& \int f\{ln(f(.|x_1))-ln(g^*(.|x_1))+ln(f_1(x_1))-ln(g_1(x_1)r(x_1))\}dx,\nonumber\\
        &=& \int f\{ln(f(.|x_1))-ln(g(.|x_1)) +ln(f_1(x_1))-ln(g_1(x_1)r(x_1))\}dx,\label{dernière}
\end{eqnarray}
as $g^*(.|x_1)=g(.|x_1)$. Since the minimum of this last equation (\ref{dernière}) is reached through the minimization of $\int f\{ln(f_1(x_1))-ln(g_1(x_1)r(x_1))\}dx=  K(f_1,g_1r)$,
then property  \ref{Phimini} necessarily implies that $f_1=g_1r$, hence $r=f_1/g_1$.\\
Finally, we have $K(f,g)-  K(f,g^*)= \int f\{ln(f_1(x_1))-ln(g_1(x_1))\}dx=K(f_1,g_1),$
which completes the demonstration of proposition \ref{lemmeHuberprop}.\\
Similarly, if we replace $f^*=fr^{-1}$ with $f$ and $g$ with $g^*$, we obtain the proposition \ref{lemmeHuber0prop}.\hfill$\Box$\\
\noindent{\bf Proof of proposition \ref{liendeminimaxi}.}
The demonstration is very similar to the one for proposition \ref{lemmeHuberprop}, save for the fact we now base our reasoning at row (\ref{dernière-1}) on 
$\int g\{ln(g^*)-ln(f)\}dx$ instead of $K(f,g^*)= \int f\{ln(f)-ln(g^*)\}dx$.\hfill$\Box$\\
\noindent{\bf Proof of proposition \ref{lemmeHuberModifprop}.}\\
Without loss of generality, we reason with $x_1$ in lieu of $\transp ax$.\\
Let us define $g^*=gr$. We remark that $g$ and $g^*$ present the same density conditionally to $x_1$.
Indeed, $g^*_1(x_1)= \int g^*(x)dx_2...dx_d= \int h(x_1)g(x)dx_2...dx_d= h(x_1)\int g(x)dx_2...dx_d= h(x_1)g_1(x_1)$.\\
We can therefore prove  this proposition.\\
First, since $f$ and $g$ are known, then, for any given function $h:x_1\mapsto h(x_1)$, the application $T$, which is defined by:

$T:g(./x_1)\frac{h(x_1)f_1(x_1)}{g_1(x_1)}\mapsto g(./x_1)f_1(x_1)$,

$T:f(./x_1)f_1(x_1)\mapsto f(./x_1)f_1(x_1)$\\
is measurable.\\
Second, the above remark implies that

$\Phi(g^*,f)=\Phi(g^*(./x_1)\frac{g_1(x_1)h(x_1)}{f_1(x_1)},f(./x_1)f_1(x_1))=\Phi(g(./x_1)\frac{g_1(x_1)h(x_1)}{f_1(x_1)},f(./x_1)f_1(x_1)).$\\
Consequently, property \ref{ExitenceDeLEntropieDesProj} page \pageref{ExitenceDeLEntropieDesProj} infers :

$\Phi(g(./x_1)\frac{g_1(x_1)h(x_1)}{f_1(x_1)},f(./x_1)f_1(x_1))\geq \Phi(T^{-1}(g(./x_1)\frac{g_1(x_1)h(x_1)}{f_1(x_1)}),T^{-1}(f(./x_1)f_1(x_1)))$

$\ \ \ \ \ \  \ \ \ \ \ \ \ \  \ \ \ \ \ \ \ \  \ \  \ \ \ \ \ \ \ \   $ $=\Phi(g(./x_1)f_1(x_1),f(./x_1)f_1(x_1))$,
by the very definition of $T$.

$\ \ \ \ \ \  \ \ \ \ \ \ \ \  \ \ \ \ \ \ \ \  \ \  \ \ \ \ \ \ \ \   $ 
$=\Phi(g\frac{f_1}{g_1},f)$,\\
which completes the proof of this proposition.\hfill$\Box$\\
\noindent{\bf Proof of lemma \ref{ChangBasis}.}
\begin{lemme} \label{ChangBasis}
We have $g(./\transp{a_{1}}x,...,\transp{a_{j}}x)=n(\transp{a_{j+1}}x,...,\transp{a_{d}}x)=f(./\transp{a_{1}}x,...,\transp{a_{j}}x)$.
\end{lemme}
\noindent Putting $A=(a_1,..,a_d)$, let us determine $f$ in basis $A$.
Let us first study the function defined by $\psi:\R^d\to\R^d$, $x\mapsto(\transp{a_1}x,..,\transp{a_d}x).$
We can immediately say that $\psi$ is continuous and since $A$ is a basis, its bijectivity is obvious.
Moreover, let us study its Jacobian.\\
By definition, it is $J_\psi(x_1,\dotsc,x_d)=
\begin{vmatrix}
\displaystyle\frac{\partial\psi_1}{\partial x_1} & \dotsb & \displaystyle\frac{\partial\psi_1}{\partial x_d}\\
\dotsb & \dotsb & \dotsb\\
\displaystyle\frac{\partial\psi_d}{\partial x_1} & \dotsb & 
\displaystyle\frac{\partial\psi_d}{\partial x_d}
\end{vmatrix}=
\begin{vmatrix}
\displaystyle a_{1,1} & \dotsb & \displaystyle a_{1,d}\\
\dotsb & \dotsb & \dotsb\\
\displaystyle a_{d,1} & \dotsb & 
\displaystyle a_{d,d}
\end{vmatrix}=|A|\not=0$ since $A$ is a basis. We can therefore infer : 
$\forall x \in\R^d,\ \exists! y\in\R^d\text{ such that }f(x)=|A|^{-1}\Psi(y),$
i.e. $\Psi$ (resp. $y$) is the expression of $f$ (resp of $x$) in  basis $A$, namely
$\Psi(y)=\tilde n(y_{j+1},...,y_{d})\tilde h(y_{1},...,y_{j})$, with $\tilde n$ and $\tilde h$ being the expressions of $n$ and $h$ in basis $A$. 
Consequently, our results in the case where the family $\{a_j\}_{1\leq j\leq d}$ is the canonical basis of $\R^d$, still hold for $\Psi$ in  basis $A$ - see section \ref{gChoice}. And then, if $\tilde g$ is the expression of $g$ in basis $A$, we have
$\tilde g(./y_1,...,y_{j})= \tilde n(y_{j+1},...,y_{d})=\Psi(./y_{1},...,y_{j})$, i.e. $g(./\transp{a_{1}}x,...,\transp{a_{j}}x)=n(\transp{a_{j+1}}x,...,\transp{a_{d}}x)=f(./\transp{a_{1}}x,...,\transp{a_{j}}x)$.\hfill$\Box$\\
\noindent{\bf Proof of lemma \ref{Base}.}
\begin{lemme} \label{Base}
Should there exist a family $(a_i)_{i=1...d}$ such that 
$f(x)=n(\transp{a_{j+1}}x,...,\transp{a_{d}}x)h(\transp{a_{1}}x,...,\transp{a_{j}}x),$
with $j<d$, with $f$, $n$ and $h$ being densities, then this family is a orthogonal basis of $\R^d$.
\end{lemme}
\noindent Using a reductio ad absurdum, we have $\int f(x)dx=1\not=+\infty=\int n(\transp{a_{j+1}}x,...,\transp{a_{d}}x)h(\transp{a_{1}}x,...,\transp{a_{j}}x)dx$. We can therefore conclude.\hfill$\Box$
$\\$\noindent{\bf Proof of proposition  \ref{QuotientDonneLoi}.}\\
Let us note first that we will prove this proposition for $k\geq2$, i.e. in the case where $g^{(k-1)}$ is not known. The initial case using the known density $g^{(0)}=g$, will be an immediate consequence from the above. \\
Moreover, going forward, to be more legible, we will use $g$ (resp. $g_n$) in lieu of $g^{(k-1)}$ (resp. $g^{(k-1)}_n$).\\
We can therefore remark that we have $f(X_i)\geq \theta_n-y_n$, $g(Y_i)\geq \theta_n-y_n$ and $g_b(\transp bY_i)\geq \theta_n-y_n$, for all $i$ and for all $b\in \R^d_*$, thanks to the uniform convergence of the kernel estimators.
Indeed, we have $f(X_i)=f(X_i)-f_n(X_i)+f_n(X_i)\geq-y_n+f_n(X_i)$, by definition of $y_n$, and then  $f(X_i)\geq-y_n+\theta_n$, by hypothesis on $f_n(X_i)$. This is also true for $g_n$ and $g_{b,n}$.\\
This entails
$\sup_{b\in\R^d_*}|\frac{1}{n}\Sigma_{i=1}^{n}\varphi'(\frac{f_{a,n}(\transp aY_i)}{g_{a,n}(\transp aY_i) }\frac{g_n(Y_i)}{f_n(Y_i)}).\frac{f_{a,n}(\transp aY_i)}{g_{a,n}(\transp aY_i) }-\int \varphi'(\frac{g(x)}{f(x)}\frac{f_b(\transp bx)}{g_b(\transp bx)}) g(x)\frac{f_a(\transp ax)}{g_a(\transp ax)} dx|\to0\ a.s.$\\
Indeed, let us remark that

$|\frac{1}{n}\Sigma_{i=1}^{n}\{\varphi'\{\frac{f_{a,n}(\transp aY_i)}{g_{a,n}(\transp aY_i) }\frac{g_n(Y_i)}{f_n(Y_i)}\}\frac{f_{a,n}(\transp aY_i)}{g_{a,n}(\transp aY_i) }\}-\int\ \varphi'(\frac{g(x)}{f(x)}\frac{f_b(\transp bx)}{g_b(\transp bx)})\ g(x)\frac{f_a(\transp ax)}{g_a(\transp ax)}\ dx|$

$=|\frac{1}{n}\Sigma_{i=1}^{n}\varphi'\{\frac{f_{a,n}(\transp aY_i)}{g_{a,n}(\transp aY_i) }\frac{g_n(Y_i)}{f_n(Y_i)}\}\frac{f_{a,n}(\transp aY_i)}{g_{a,n}(\transp aY_i) }-\frac{1}{n}\Sigma_{i=1}^{n}\varphi'\{\frac{f_{a}(\transp aY_i)}{g_{a}(\transp aY_i) }\frac{g(Y_i)}{f(Y_i)}\}\frac{f_{a}(\transp aY_i)}{g_{a}(\transp aY_i) }$

$\ \ \ \ \ $ $+\frac{1}{n}\Sigma_{i=1}^{n}\varphi'\{\frac{f_{a}(\transp aY_i)}{g_{a}(\transp aY_i) }\frac{g(Y_i)}{f(Y_i)}\}\frac{f_{a}(\transp aY_i)}{g_{a}(\transp aY_i) }-\int\ \varphi'(\frac{g(x)}{f(x)}\frac{f_b(\transp bx)}{g_b(\transp bx)})\ g(x)\frac{f_a(\transp ax)}{g_a(\transp ax)}\ dx|$

$\leq|\frac{1}{n}\Sigma_{i=1}^{n}\varphi'\{\frac{f_{a,n}(\transp aY_i)}{g_{a,n}(\transp aY_i) }\frac{g_n(Y_i)}{f_n(Y_i)}\}\frac{f_{a,n}(\transp aY_i)}{g_{a,n}(\transp aY_i) }-\frac{1}{n}\Sigma_{i=1}^{n}\varphi'\{\frac{f_{a}(\transp aY_i)}{g_{a}(\transp aY_i) }\frac{g(Y_i)}{f(Y_i)}\}\frac{f_{a}(\transp aY_i)}{g_{a}(\transp aY_i) }|$

$\ \ \ \ \ $ $+|\frac{1}{n}\Sigma_{i=1}^{n}\varphi'\{\frac{f_{a}(\transp aY_i)}{g_{a}(\transp aY_i) }\frac{g(Y_i)}{f(Y_i)}\}\frac{f_{a}(\transp aY_i)}{g_{a}(\transp aY_i) }-\int\ \varphi'(\frac{g(x)}{f(x)}\frac{f_b(\transp bx)}{g_b(\transp bx)})\ g(x)\frac{f_a(\transp ax)}{g_a(\transp ax)}\ dx|$\\
Moreover, since $\int|\varphi'(\frac{g(x)}{f(x)}\frac{f_b(\transp bx)}{g_b(\transp bx)})\ g(x)\frac{f_a(\transp ax)}{g_a(\transp ax)}|dx< \infty$, as implied by lemma \ref{ExitenceDeLEntropieDesProj}, and since we assumed $g$ such that $\Phi(g,f)<\infty$ and $\Phi(f,g)<\infty$ and since $b\in \Theta^{\Phi} $, the law of large numbers enables us to state that $|\frac{1}{n}\Sigma_{i=1}^{n}\varphi'\{\frac{f_{a}(\transp aY_i)}{g_{a}(\transp aY_i) }\frac{g(Y_i)}{f(Y_i)}\}\frac{f_{a}(\transp aY_i)}{g_{a}(\transp aY_i) }-\int\ \varphi'(\frac{g(x)}{f(x)}\frac{f_b(\transp bx)}{g_b(\transp bx)})\ g(x)\frac{f_a(\transp ax)}{g_a(\transp ax)}\ dx|\to 0\ a.s.$\\
Furthermore, $|\frac{1}{n}\Sigma_{i=1}^{n}\varphi'\{\frac{f_{a,n}(\transp aY_i)}{g_{a,n}(\transp aY_i) }\frac{g_n(Y_i)}{f_n(Y_i)}\}\frac{f_{a,n}(\transp aY_i)}{g_{a,n}(\transp aY_i) }-\frac{1}{n}\Sigma_{i=1}^{n}\varphi'\{\frac{f_{a}(\transp aY_i)}{g_{a}(\transp aY_i) }\frac{g(Y_i)}{f(Y_i)}\}\frac{f_{a}(\transp aY_i)}{g_{a}(\transp aY_i) }|$

$\ \ \ \ \ \ \ \ \ \ \ \ \ $ $\leq \frac{1}{n}\Sigma_{i=1}^{n}|\varphi'\{\frac{f_{a,n}(\transp aY_i)}{g_{a,n}(\transp aY_i) }\frac{g_n(Y_i)}{f_n(Y_i)}\}\frac{f_{a,n}(\transp aY_i)}{g_{a,n}(\transp aY_i) }-\varphi'\{\frac{f_{a}(\transp aY_i)}{g_{a}(\transp aY_i) }\frac{g(Y_i)}{f(Y_i)}\}\frac{f_{a}(\transp aY_i)}{g_{a}(\transp aY_i) }|$\\
and $|\varphi'\{\frac{f_{a,n}(\transp aY_i)}{g_{a,n}(\transp aY_i) }\frac{g_n(Y_i)}{f_n(Y_i)}\}\frac{f_{a,n}(\transp aY_i)}{g_{a,n}(\transp aY_i) }-\varphi'\{\frac{f_{a}(\transp aY_i)}{g_{a}(\transp aY_i) }\frac{g(Y_i)}{f(Y_i)}\}\frac{f_{a}(\transp aY_i)}{g_{a}(\transp aY_i) }|\to 0$
as a result of the hypotheses intially introduced on $\theta_n.$
Consequently, $\frac{1}{n}\Sigma_{i=1}^{n}|\varphi'\{\frac{f_{a,n}(\transp aY_i)}{g_{a,n}(\transp aY_i) }\frac{g_n(Y_i)}{f_n(Y_i)}\}\frac{f_{a,n}(\transp aY_i)}{g_{a,n}(\transp aY_i) }-\varphi'\{\frac{f_{a}(\transp aY_i)}{g_{a}(\transp aY_i) }\frac{g(Y_i)}{f(Y_i)}\}\frac{f_{a}(\transp aY_i)}{g_{a}(\transp aY_i) }|\to 0$, as it is a  Cesàro mean. This enables us to conclude. Similarly, we obtain

$\sup_{b\in\R^d_*}|\frac{1}{n}\Sigma_{i=1}^n\varphi^*\{\varphi'\{\frac{f_{a,n}(\transp aX_i)}{g_{a,n}(\transp aX_i) }\frac{g_n(X_i)}{f_n(X_i)}\}\}-\ \int\ \varphi^*(\varphi'(\frac{g(x)}{f(x)}\frac{f_b(\transp bx)}{g_b(\transp bx)}))f(x)dx|\to0\ a.s.$\hfill$\Box$\\
\noindent{\bf Proof of lemma \ref{compacité-1}.} 
By definition of the closure of a set, we have
\begin{lemme} \label{compacité-1}
The set $\Gamma_c$ is closed in $L^1$ for the topology of the uniform convergence.
\end{lemme}
\noindent{\bf Proof of lemma \ref{compacité}.} 
Since $\Phi$ is greater than the $L^1$ distance, we have
\begin{lemme} \label{compacité}
For all $c>0$, we have $\Gamma_c\subset \overline B_{L^1}(f,c),$ where $B_{L^1}(f,c)=\{p\in L^1;\|f-p\|_1\leq c\}$.
\end{lemme}
\noindent{\bf Proof of lemma \ref{compacité+1}.} 
The definition of the closure of a set and lemma \ref{aleph} (see page \pageref{aleph}) imply
\begin{lemme} \label{compacité+1}
 $G$ is closed in $L^1$ for the topology of the uniform convergence.
\end{lemme}
\noindent{\bf Proof of lemma \ref{toattain}.}
\begin{lemme}\label{toattain}
$\inf_{a\in\R^d_*}  \Phi(g^*,f)$ is reached when the $\Phi$-divergence is greater than the $L^1$ distance as well as the $L^2$ distance.
\end{lemme}
\begin{proof}
Indeed, let $G$ be $\{g\frac{f_a}{g_a};\ a\in\R^d_*\}$ and $\Gamma_c$ be $\Gamma_c=\{p;\   K(p,f)\leq c\}$ for all $c$>0. From lemmas \ref{compacité-1}, \ref{compacité} and \ref{compacité+1} (see page \pageref{compacité}), we get $\Gamma_c\cap G$ is a compact for the topology of the uniform convergence, if $\Gamma_c\cap G$ is not empty.
Hence, and since  property \ref{K-SCI} (see page \pageref{K-SCI}) implies that  $Q\mapsto   \Phi(Q,P)$ is lower semi-continuous in $L^1$ for the topology of the uniform convergence, then the infimum is reached in $L^1$.
(Taking for example  $c=\Phi(g,f),$ $\Omega$  is necessarily not empty because we always have $\Phi(g\frac{f_a}{g_a},f)\leq \Phi(g,f)$).
Moreover, when the $\Phi-$divergence is greater than the $L^2$ distance, the very definition of the $L^2$ space enables us to provide the same proof as for the $L^1$ distance.
\end{proof}
\noindent{\bf Proof of lemma \ref{TrucBidule}.}
\begin {lemme}\label{TrucBidule}
For any $p\leq d$, we have $f^{(p-1)}_{a_p}=f_{a_p}$ - see Huber's analytic method -, $g^{(p-1)}_{a_p}=g_{a_p}$ -  see Huber's synthetic method - and $g^{(p-1)}_{a_p}=g_{a_p}$ - see our algorithm.
\end{lemme}
\begin{proof}
As it is equivalent to prove either our algorithm or Huber's, we will only develop here the proof for our algorithm.
Assuming, without any loss of generality, that the $a_i$, $i=1,..,p$, are the vectors of the canonical basis, since 
$g^{(p-1)}(x)=g(x)\frac{f_1(x_1)}{g_1(x_1)}\frac{f_2(x_2)}{g_2(x_2)}...\frac{f_{p-1}(x_{p-1})}{g_{p-1}(x_{p-1})}$ we derive immediately that $g^{(p-1)}_{p}=g_{p}$. We note that it is sufficient to operate a change in basis on the $a_i$ to obtain the general case.
\end{proof}
\noindent{\bf Proof of lemma \ref{imFree}.}
\begin {lemme}\label{imFree}
If there exits $p$, $p\leq d$, such that $\Phi(g^{(p)},f)=0$, then the family  of $(a_i)_{i=1,..,p}$ - derived from the construction of $g^{(p)}$ - is free and orthogonal.
\end{lemme}
\begin{proof}
Without any loss of generality, let us assume that $p=2$ and that the $a_i$ are the vectors of the canonical basis. Using a reductio ad absurdum with the hypotheses  $a_1=(1,0,...,0)$ and that $a_2=(\alpha,0,...,0)$, where  $\alpha\in\R$, we get $g^{(1)}(x)=g(x_2,..,x_d/x_1)f_1(x_1)$ and $f=g^{(2)}(x)=g(x_2,..,x_d/x_1)f_1(x_1)\frac{f_{\alpha a_1}(\alpha x_1)}{[g^{(1)}]_{\alpha a_1}(\alpha x_1)}$. Hence $f(x_2,..,x_d/x_1)=g(x_2,..,x_d/x_1)\frac
{f_{\alpha a_1}(\alpha x_1)}
{[g^{(1)}]_{\alpha a_1}(\alpha x_1)}.$\\
It consequently implies that $f_{\alpha a_1}(\alpha x_1)=[g^{(1)}]_{\alpha a_1}(\alpha x_1)$ since\\
 $1=\int f(x_2,..,x_d/x_1)dx_2...dx_d=\int g(x_2,..,x_d/x_1)dx_2...dx_d\frac
{f_{\alpha a_1}(\alpha x_1)}
{[g^{(1)}]_{\alpha a_1}(\alpha x_1)}=\frac
{f_{\alpha a_1}(\alpha x_1)}
{[g^{(1)}]_{\alpha a_1}(\alpha x_1)}$.\\
Therefore, $g^{(2)}=g^{(1)}$, i.e. $p=1$ which leads to a contradiction. Hence, the family is free.\\
Moreover, using a reductio ad absurdum we get the orthogonality. Indeed, we have \\
$\int f(x)dx=1\not=+\infty=\int n(\transp{a_{j+1}}x,...,\transp{a_{d}}x)h(\transp{a_{1}}x,...,\transp{a_{j}}x)dx$. The use of the same argument as in the proof of lemma \ref{Base}, enables us to infer the orthogonality of $(a_i)_{i=1,..,p}$.
\end{proof}
\noindent{\bf Proof of lemma \ref{OrthoOfVect}.}
\begin {lemme}\label{OrthoOfVect}
If there exits $p$, $p\leq d$, such that $\Phi(g^{(p)},f)=0$, where $g^{(p)}$ is built from the free and orthogonal family $a_{1}$,...,$a_{j}$, then, there exists a free and orthogonal family $(b_k)_{k=j+1,...,d}$ of vectors of $\R^d_*$, such that 
$g^{(p)}(x)=g(\transp b_{j+1}x,...,\transp b_{d}x/\transp a_{1}x,...,\transp a_{j}x)f_{a_{1}}(\transp a_{1}x)...f_{a_{j}}(\transp a_{j}x)$\\
and such that $\R^d=Vect\{a_i\}\stackrel{\perp}{\oplus}Vect\{b_k\}$.
\end{lemme}
\begin{proof}
Through the incomplete basis theorem and similarly as in lemma \ref{imFree}, we obtain the result thanks to the Fubini's theorem.
\end{proof}
\noindent{\bf Proof of lemma \ref{KernRate}.}
\begin{lemme}\label{KernRate}
For any continuous density $f$, we have
$y_m=|f_m(x)-f(x)|=O_\PP(m^{-\frac{2}{4+d}})$.
\end{lemme}
Defining $b_m(x)$ as  $b_m(x)=|E(f_m(x))-f(x)|$, we have $y_m\leq |f_m(x)-E(f_m(x))|+b_m(x)$. Moreover, 
from page 150 of \cite{MR1191168}, we derive that $b_m(x)=O_\PP(\Sigma_{j=1}^dh_j^2)$ where $h_j=O_\PP(m^{-\frac{1}{4+d}})$. Then, we obtain $b_m(x)=O_\PP(m^{-\frac{2}{4+d}})$. Finally, since the central limit theorem rate  is $O_\PP(m^{-\frac{1}{2}})$, we infer that $y_m\leq O_\PP(m^{-\frac{1}{2}})+O_\PP(m^{-\frac{2}{4+d}})=O_\PP(m^{-\frac{2}{4+d}})$.\hfill$\Box$\\
\noindent{\bf Proof of proposition \ref{KernelpConv2}.}
Proposition \ref{KernelpConv2} comes immediately from proposition \ref{QuotientDonneLoi} page \pageref{QuotientDonneLoi} and lemma \ref{pConv2} page \pageref{pConv2}.\hfill$\Box$\\
\noindent{\bf Proof of proposition \ref{HuberApp}.}
Let us first show by induction the following assertion

$\cP(k)=\{g^{(k)}\text{ allows a deconvolution }g^{(k)}=\overline g^{(k)}*\phi\}$\\
Initialisation :
For $k=0$, we get the result since $g=g^{(0)}$ is elliptic.\\
Going from $k$ to $k+1$ :
Let us assume $\cP(k)$ is true, we then show that $\cP(k+1)$. \\
Since the family of $a_i$, $i\leq k+1$ is free - see lemma \ref{imFree} - then, we define $B$ as the basis of $\R^d$ such that its $k+1$ first vectors are the $a_i$, $i\leq k+1$ - see the incomplete basis theorem for its existence.\\
Thus, in $B$ and using the same procedure to prove lemma \ref{ChangBasis} page \pageref{ChangBasis}, we have \\$\overline g^{(k)}(x)=\overline g^{(k)}(./x_{k+1})\overline g^{(k)}_{k+1}(x_{k+1})$. Consequently, the very definition of the convolution product, the Fubini's theorem and the hypothesis made on the Elliptical family imply that \\  $g^{(k)}(x)=g^{(k)}(./x_{k+1})g^{(k)}_{k+1}(x_{k+1})$ with $g^{(k)}(./x_{k+1})=\overline g^{(k)}(./x_{k+1})*E_{d-1}(0,\sigma^2I_{d-1},\xi_{d-1})$ and with $g^{(k)}_{k+1}(x_{k+1})=\overline g^{(k)}_{k+1}(x_{k+1})*E_{1}(0,\sigma^2,\xi_{1})$.
Finally, replacing $g^{(k)}_{k+1}$ with $f_{k+1}=\overline f_{k+1}*E_{1}(0,\sigma^2,\xi_{1})$, we conclude this induction with $g^{(k+1)}=g^{(k)}(./x_{k+1})f_{k+1}(x_{k+1})$.\\
Now, let us consider $\psi$ (rep. $\overline \psi$, $\psi^{(k)}$, $\overline \psi^{(k)}$) the characteristic function of $f$ (resp. $\overline f$, $g^{(k)}$, $\overline g^{(k)}$). We then have
$\psi(s)=\overline \psi(s)\Psi(\frac{1}{2}\sigma^2|s|^2)$ and $\psi^{(k)}(s)=\overline \psi^{(k)}(s)\Psi(\frac{1}{2}\sigma^2|s|^2)$. Hence, $\psi$ and $\psi^{(k)}$ are less or equal to $\Psi(\frac{1}{2}\sigma^2|s|^2)$ which is integrable by hypothesis, i.e. $\psi$ and $\psi^{(k)}$ are absolutely integrable. We then obtain
$g^{(k)}(x)=(2\pi)^{-d}\int \psi^{(k)}(s)e^{-i\transp sx}ds$
and $f(x)=(2\pi)^{-d}\int \psi(s)e^{-i\transp sx}ds$.\\
Moreover, since the sequence $(\psi^{(k)})$ uniformly converges and since $\psi$ and $\psi^{(k)}$ are less or equal to $\Psi(\frac{1}{2}\sigma^2|s|^2)$, then the dominated convergence theorem implies that\\
$\lim_k|f(x)-g^{(k)}(x)|\leq (2\pi)^{-d}\int \lim_k|\psi(s)-\psi^{(k)}(s)|ds=0$ a.s.
i.e. $\lim_ksup_x|f(x)-g^{(k)}(x)|=0$ a.s.\\
Finally, since, by hypothesis, $(2\pi)^{-d}\int |\psi(s)-\psi^{(k)}(s)|ds\leq 2(2\pi)^{-d}\int\Psi(\frac{1}{2}\sigma^2|s|^2)ds<\infty$, then the above limit and the dominated convergence theorem imply that $\lim_k\int|f(x)-g^{(k)}(x)|dx=0.$
\hfill$\Box$\\
\noindent{\bf Proof of corollary \ref{FromSection14Huber-1}.}
Through the dominated convergence theorem and through theorem \ref{limnk}, we get the result using a reductio ad absurdum.\hfill$\Box$\\
\noindent{\bf Proof of lemma \ref{FromSection14Huber}.}
\begin{lemme}\label{FromSection14Huber}
Let consider the sequence $(a_i)$ defined in (\ref{VraiDefOfAK}) page \pageref{VraiDefOfAK}.\\
We then have $\lim_n\lim_kK(\check g^{(k)}_n\frac{f_{{a_k},n}}{[\check g^{(k)}]_{{a_k},n}},f_n)= 0$ a.s.
\end{lemme}
\begin{proof}
Trough the relationship (\ref{VraiDefOfAK}) and through remark \ref{criteria-H} page \pageref{criteria-H} as well as the additive relation of proposition \ref{lemmeHuber0prop}, we can say that
$0\leq..\leq K(g^{(\infty)},f)\leq..\leq K(g^{(k)},f)\leq..\leq K(g,f)$,
where $g^{(\infty)}=\lim_kg^{(k)}$ which is a density by construction. And through proposition \ref{cvl}, we obtain that $K(g^{(\infty)},f)=0$, i.e.

$0=K(g^{(\infty)},f)\leq\ldots\leq K(g^{(k)},f)\leq\ldots\leq K(g,f)$, (*).\\
Moreover, let $(g^{(k)}_n)_k$ be the sequence of densities such that $g^{(k)}_n$ is the kernel estimate of $g^{(k)}$. Since we derive from remark \ref{GkBor} page \pageref{GkBor} an integrable upper bound of $g^{(k)}_n$, for all $k$, which is greater than $f$ - see also the definition of $\varphi$ in the proof of theorem \ref{limnk} -, then the dominated convergence theorem implies that, for any $k$, $\lim_nK(g^{(k)}_n,f_n)=K(g^{(k)},f)$, i.e., from a certain given rank $n_0$, we have $0\leq..\leq K(g^{(\infty)}_n,f_n)\leq ..\leq K(g^{(k)}_n,f_n)\leq..\leq K(g_n,f_n)$, (**).\\
Consequently, through lemma \ref{Ineq***} page \pageref{Ineq***}, there exists a $k$ such that

$0\leq..\leq K(\Psi^{(\infty)}_{n,k},f_n)\leq..\leq K(g^{(\infty)}_n,f_n)\leq ..\leq K(\Psi^{(\infty)}_{n,k-1},f_n)\leq..\leq K(g_n,f_n)$, (***)\\
where $\Psi^{(\infty)}_{n,k}$ is a density such that $\Psi^{(\infty)}_{n,k}=\lim_kg^{(k)}_n$.\\
Finally, through the dominated convergence theorem and taking the limit as $n$ in (***) we get

$0=K(g^{(\infty)},f) = \lim_nK(g^{(\infty)}_n,f_n)\geq \lim_nK(\Psi^{(\infty)}_{n,k},f_n)\geq 0$.\\
The dominated convergence theorem enables us to conclude:

$0=\lim_nK(\Psi^{(\infty)}_{n,k},f_n)=\lim_n\lim_kK(g^{(k)}_n,f_n)$.
\end{proof}
\noindent{\bf Proof of lemma \ref{Ineq***}.}
\begin{lemme}\label{Ineq***}
With the notation of the proof of lemma \ref{FromSection14Huber}, we have 

$0\leq..\leq K(\Psi^{(\infty)}_{n,k},f_n)\leq..\leq K(g^{(\infty)}_n,f_n)\leq ..\leq K(\Psi^{(\infty)}_{n,k-1},f_n)\leq..\leq K(g_n,f_n)$, (***)
\end{lemme}
\begin{proof}
First, as explained in section \ref{a1solve4}, we have $K(f^{(k)},g)-K(f^{(k+1)},g)=K(f^{(k)}_{a_{k+1}},g_{a_{k+1}})$. Moreover, through remark \ref{criteria-H} page \pageref{criteria-H}, we also derive that $K(f^{(k)},g)=K(g^{(k)},f)$. Then, $K(f^{(k)}_{a_{k+1}},g_{a_{k+1}})$ is the decreasing step of the relative entropies in (*) and leading to $0=K(g^{(\infty)},f)$. Similarly, the very construction of (**), implies that $K(f^{(k)}_{a_{k+1},n},g_{a_{k+1},n})$ is the decreasing step of the relative entropies in (**) and leading to  $K(g^{(\infty)}_n,f_n)$.\\
Second, through the conclusion of the section \ref{a1solve4} and lemma 14.2 of Huber's article, we obtain that $K(f^{(k)}_{a_{k+1},n},g_{a_{k+1},n})$ converges - in decreasing and in $k$ - towards a positive function of $n$ - that we will call $\xi_n$.\\
Third, the convergence of $(g^{(k)})_k$ - see proposition \ref{cvl} - implies that, for any given $n$, the sequence $(K(g^{(k)}_n,f_n))_k$ is not finite. Then, through relationship $(**)$, there exists a $k$ such that $0<K(g^{(k-1)}_n,f_n)-K(g^{(\infty)}_n,f_n)<\xi_n$.\\
Thus, since $Q\mapsto K(Q,P)$ is l.s.c. - see property \ref{K-SCI} page \pageref{K-SCI} - relationship (**) implies (***).
\end{proof}
\noindent{\bf Proof of theorem \ref{KernelKRessultatPricipal}.}
First, by the very definition of the kernel estimator $\check g^{(0)}_n=g_n$ converges towards $g$. Moreover, the continuity of $a\mapsto f_{a,n}$ and $a\mapsto g_{a,n}$ and proposition \ref{KernelpConv2} imply that $\check g^{(1)}_n=\check g^{(0)}_n\frac{f_{a,n}}{\check g^{(0)}_{a,n}}$ converges towards $g^{(1)}$. 
Finally, since, for any $k$, $\check g^{(k)}_n=\check g^{(k-1)}_n\frac{f_{\check a_k,n}}{\check g^{(k-1)}_{\check a_k,n}}$, we conclude by an immediat induction.\hfill$\Box$\\
\noindent{\bf Proof of theorem \ref{Superdiffconv}.}\\
\noindent{\bf relationship (\ref{diffconv1}).}
Let us consider $\Psi_j=\{\frac{f_{\check{a_j}}(\transp{\check{a_j}}x)}{[\check g^{(j-1)}]_{\check{a_j}}(\transp{\check{a_j}}x)}-\frac{f_{a_j}(\transp{a_j}x)}{[g^{(j-1)}]_{a_j}(\transp{a_j}x)}\}$.
Since $f$ and $g$ are bounded, it is easy to prove  that from a certain rank, we get, for any $x$ given in $\R^d$

$|\Psi_j| \leq max(\frac{1}{[\check g^{(j-1)}]_{\check{a_j}}(\transp{\check{a_j}}x)},\frac{1}{[ g^{(j-1)}]_{a_j}(\transp{a_j}x)})|f_{\check{a_j}}(\transp{\check{a_j}}x)-f_{a_j}(\transp{a_j}x)|$.
\begin{remarque}
First, based on what we stated earlier, for any given $x$ and from a certain rank, 
there is a constant $R$>0 independent from $n$, such that  

$max(\frac{1}{[\check g^{(j-1)}]_{\check{a_j}}(\transp{\check{a_j}}x)},\frac{1}{[ g^{(j-1)}]_{a_j}(\transp{a_j}x)})\leq R=R(x)=O(1).$\\
Second, since  $\check a_k$ is an $M-$estimator of $a_k$, its convergence rate is $O_{\PP}(n^{-1/2})$.
\end{remarque}
Thus using simple functions, we infer an upper and lower bound for $f_{\check{a_j}}$ and for $f_{a_j}$. We therefore reach the following conclusion:
\begin{equation}\label{PsiO}
|\Psi_j|\leq O_{\PP}(n^{-1/2}).
\end{equation}
We finally obtain

$|\Pi_{j=1}^k\frac{f_{\check{a_j}}(\transp{\check{a_j}}x)}{[\check g^{(j-1)}]_{\check{a_j}}(\transp{\check{a_j}}x)}-\Pi_{j=1}^k\frac{f_{a_j}(\transp{a_j}x)}{[g^{(j-1)}]_{a_j}(\transp{a_j}x)}|=\Pi_{j=1}^k\frac{f_{a_j}(\transp{a_j}x)}{[g^{(j-1)}]_{a_j}(\transp{a_j}x)}|\Pi_{j=1}^k\frac{f_{\check{a_j}}(\transp{\check{a_j}}x)}{[\check g^{(j-1)}]_{\check{a_j}}(\transp{\check{a_j}}x)}\frac{[g^{(j-1)}]_{a_j}(\transp{a_j}x)}{f_{a_j}(\transp{a_j}x)}-1|$.\\
Based on relationship (\ref{PsiO}), the expression   $\frac{f_{\check{a_j}}(\transp{\check{a_j}}x)}{[\check g^{(j-1)}]_{\check{a_j}}(\transp{\check{a_j}}x)}\frac{[g^{(j-1)}]_{a_j}(\transp{a_j}x)}{f_{a_j}(\transp{a_j}x)}$ tends towards 1 at a rate of $O_{\PP}(n^{-1/2})$ for all $j$.
Consequently, $\Pi_{j=1}^k\frac{f_{\check{a_j}}(\transp{\check{a_j}}x)}{[\check g^{(j-1)}]_{\check{a_j}}(\transp{\check{a_j}}x)}\frac{[g^{(j-1)}]_{a_j}(\transp{a_j}x)}{f_{a_j}(\transp{a_j}x)}$ tends towards 1 at a rate of $O_{\PP}(n^{-1/2})$. Thus from a certain rank, we get

$|\Pi_{j=1}^k\frac{f_{\check{a_j}}(\transp{\check{a_j}}x)}{[\check g^{(j-1)}]_{\check{a_j}}(\transp{\check{a_j}}x)}-\Pi_{j=1}^k\frac{f_{a_j}(\transp{a_j}x)}{[g^{(j-1)}]_{a_j}(\transp{a_j}x)}|=O_{\PP}(n^{-1/2})O_{\PP}(1)=O_{\PP}(n^{-1/2})$.\\
In conclusion, we obtain 
$|\check g^{(k)}(x)-g^{(k)}(x)|=  g(x)|\Pi_{j=1}^k\frac{f_{\check{a_j}}(\transp{\check{a_j}}x)}{[\check g^{(j-1)}]_{\check{a_j}}(\transp{\check{a_j}}x)}-\Pi_{j=1}^k\frac{f_{a_j}(\transp{a_j}x)}{[g^{(j-1)}]_{a_j}(\transp{a_j}x)}|\leq  O_{\PP}(n^{-1/2})$.\\
\noindent{\bf relationship (\ref{diffconv2}).}
The relationship \ref{diffconv1} of theorem \ref{Superdiffconv} implies that $|\frac{\check g^{(k)}(x)}{g^{(k)}(x)}-1|=O_{\PP}(n^{-1/2})$ because, for any given $x$, $g^{(k)}(x)|\frac{\check g^{(k)}(x)}{g^{(k)}(x)}-1|=|\check g^{(k)}(x)-g^{(k)}(x)|$. Consequently, there exists a smooth function $C$ of $\R^d$ in $\R^+$ such that

$\lim_{n\to\infty}n^{-1/2}C(x)=0$ and $|\frac{\check g^{(k)}(x)}{g^{(k)}(x)}-1|\leq n^{-1/2}C(x)$, for any $x$.\\
We then have
$\int |\check g^{(k)}(x)-g^{(k)}(x)|dx=\int g^{(k)}(x)|\frac{\check g^{(k)}(x)}{g^{(k)}(x)}-1|dx
\leq\int g^{(k)}(x)C(x)n^{-1/2}dx$.\\
Moreover, $\sup_{x\in\R^d}|\check g^{(k)}(x)-g^{(k)}(x)|=\sup_{x\in\R^d}g^{(k)}(x)|\frac{\check g^{(k)}(x)}{g^{(k)}(x)}-1|$

$\ \ \ \ \ \ \ \ \ \ \ \ \ \ \ \  \ \ \ \ \ \ \ \ \ \ \ \ \ \ \ \ \ \ \ \ \ \ \ \ \ \ $ $=\sup_{x\in\R^d}g^{(k)}(x)C(x)n^{-1/2}\to0\ a.s.$, by theorem \ref{KRessultatPricipal}.\\
This implies that $\sup_{x\in\R^d}g^{(k)}(x)C(x)<\infty\ a.s.$, i.e.  $\sup_{x\in\R^d}C(x)<\infty\ a.s.$ since $g^{(k)}$ has been assumed to be positive and bounded - see remark \ref{GkBor}.\\
Thus, $\int g^{(k)}(x)C(x)dx\leq \sup C.\int g^{(k)}(x)dx=\sup C<\infty$ since $g^{(k)}$ is a density, therefore we can conclude $\int |\check g^{(k)}(x)-g^{(k)}(x)|dx\leq \sup C.n^{-1/2}=O_{\PP}(n^{-1/2}).$\hfill$\Box$\\
\noindent{\bf relationship (\ref{diffconv3}).}
We have

$K(\check g^{(k)},f)-K( g^{(k)},f) = \int f (\varphi(\frac{\check g^{(k)}}{f})-\varphi(\frac{g^{(k)}}{f}))dx \leq \int f\ S|\frac{\check g^{(k)}}{f}-\frac{g^{(k)}}{f}|dx=S\int|\check g^{(k)}-g^{(k)}|dx$\\
with the line before last being derived from theorem  \ref{azeIII4} page \pageref{azeIII4} and where $\varphi : x\mapsto xln(x)-x+1$ is a convex function and where $S>0$. 
We get the same expression as the one found in our Proof of Relationship (\ref{diffconv2}) section, we then obtain  $K(\check g^{(k)},f)-K( g^{(k)},f) \leq O_{\PP}(n^{-1/2})$. Similarly, we get $K( g^{(k)},f)-K(\check g^{(k)},f)\leq O_{\PP}(n^{-1/2})$. We can therefore conclude.
\hfill$\Box$\\
\noindent{\bf Proof of lemma \ref{n-FunctOf-m}.}
\begin{lemme}\label{n-FunctOf-m}
We keep the notations introduced in Appendix \ref{truncSample}. It holds $n=O(m^{\frac 1 2})$.
\end{lemme}
\begin{proof}
Let $N$ be the random variable such that\\ $N=\Sigma_{j=1}^m{\bf 1}_{\{f_m(X_j)\geq \theta_m,\ g(Y_j)\geq \theta_m\}}$. Since the events $\{f_m(X_j)\geq \theta_m\}$ and $\{ g(Y_j)\geq \theta_m\}$ are independent from one another and since $\{g(Y_j)\geq \theta_m\}\subset\{ g_m(Y_j)\geq -y_m+\theta_m\}$, we can say that

$n=m.\PP(f_m(X_j)\geq \theta_m,\ g(Y_j)\geq \theta_m)\leq m.\PP(f_m(X_j)\geq \theta_m).\PP(g_m(Y_j)\geq -y_m+\theta_m)$.\\
Consequently, let us study $\PP(f_m(X_i)\geq \theta_m)$.
Let $(\xi_i)_{i=1\ldots m}$ be the sequence such that, for any $i$ and any $x$ in $\R^d$,
$\xi_i(x)=\Pi_{l=1}^d\frac{1}{(2\pi)^{1/2}h_l}e^{-\frac{1}{2}(\frac{x_l-X_{il}}{h_l})^2}-\int \Pi_{l=1}^d\frac{1}{(2\pi)^{1/2}h_l}e^{-\frac{1}{2}(\frac{x_l-X_{il}}{h_l})^2}\ f(x)dx.$
Hence, for any given $j$ and conditionally to $X_1,$ $\ldots,$ $X_{j-1}$, $X_{j+1},$ $\ldots,$ $X_m$, the variables $(\xi_i(X_j))_{i=1\ldots m}^{i\not=j}$ are i.i.d. and centered, have same second moment, and are such that 

$|\xi_i(X_j)|\leq \Pi_{l=1}^d\frac{1}{(2\pi)^{1/2}h_l}+\Pi_{l=1}^d\frac{1}{(2\pi)^{1/2}h_l}\int  |f(x)|dx=2.(2\pi)^{-d/2}\Pi_{l=1}^dh_l^{-1}$ since $\sup_xe^{-\frac{1}{2}x^2}\leq 1$.\\
Moreover, noting that
$f_m(x)=\frac{1}{m}\Sigma_{i=1}^m\xi_i(x)+(2\pi)^{-d/2}\frac{1}{m}\Sigma_{i=1}^m\Pi_{l=1}^dh_l^{-1}\int e^{-\frac{1}{2}(\frac{x_l-X_{il}}{h_l})^2}\ f(x)dx$, \\
we have $f_m(X_j)\geq \theta_m \Leftrightarrow\frac{1}{m}\Sigma_{i=1}^m\xi_i(X_j)+(2\pi)^{-d/2}\frac{1}{m}\Sigma_{i=1}^m\Pi_{l=1}^dh_l^{-1}\int e^{-\frac{1}{2}(\frac{x_l-X_{il}}{h_l})^2}\ f(x)dx\geq \theta_m$

$\ \ \ \ \ $ $\Leftrightarrow\frac{1}{m-1}\Sigma_{\stackrel{i=1} {i\not=j}}^m\xi_i(X_j)\geq(\theta_m-(2\pi)^{-d/2}\frac{1}{m}\Sigma_{i=1}^m\Pi_{l=1}^dh_l^{-1}\int e^{-\frac{1}{2}(\frac{x_l-X_{il}}{h_l})^2}\ f(x)dx-\frac{1}{m}\xi_j(X_j))\frac{m}{m-1}$\\
with $\xi_j(X_j)=0$. Then, defining $t$ (resp. $\varepsilon$) as $t=2.(2\pi)^{-d/2}\Pi_{l=1}^dh_l^{-1}$ (resp. \\ $\varepsilon=(\theta_m-(2\pi)^{-d/2}\Pi_{l=1}^dh_l^{-1}\frac{1}{m}\Sigma_{i=1}^m\Pi_{l=1}^d\int e^{-\frac{1}{2}(\frac{x_l-X_{il}}{h_l})^2}\ f(x)dx)\frac{m}{m-1}$), the Bennet's inequality -\cite{MR0851019} page 160- implies that 
$\PP(\frac{1}{m-1}\Sigma_{\stackrel{i=1}{i\not=j}}^m\xi_i(X_j)\geq\varepsilon/\text{$X_1,$ $\ldots,$ $X_{j-1}$, $X_{j+1},$ $\ldots,$ $X_m$})\leq2.exp(-\frac{(m-1)\varepsilon^2}{4t^2}).$\\
Finally, since the $X_i$ are i.i.d. and since $\int (\int\Pi_{l=1}^d e^{-\frac{1}{2}(\frac{x_l-y_{l}}{h_l})^2}\ f(x)dx)f(y)dy<1$, then the law of large numbers implies that 
$\frac{1}{m}\Sigma_{i=1}^m\int\Pi_{l=1}^d e^{-\frac{1}{2}(\frac{x_l-X_{il}}{h_l})^2}\ f(x)dx\to_m\int \int\Pi_{l=1}^d e^{-\frac{1}{2}(\frac{x_l-y_{l}}{h_l})^2} f(x)f(y)dxdy$ a.s.
Consequently, since $0<\nu<\frac{1}{4+d}$ - see remark \ref{Scott} - and since $e^{-x}\leq x^{-\frac{1}{2}}$ when $x>0$, we obtain, after calculation, that, from a certain rank, $exp(-\frac{(m-1)\varepsilon^2}{4t^2})=O(m^{-\frac 1 4})$, i.e., from a certain rank, $\PP(f_m(Y_j)\geq \theta_m)=O(m^{-\frac 1 4})$. Similarly, we infer $\PP(g(Y_j)\geq \theta_m)=O(m^{-\frac 1 4})$. In conclusion, we can say that 
$n=m.\PP(f_m(X_j)\geq \theta_m).\PP(g_m(Y_j)\geq \theta_m)=O(m^{\frac 1 2})$.
Similarly, we derive the same result as above for any step of our method.
\end{proof}
\noindent{\bf Proof of theorem \ref{KernelSuperdiffconv}.}
First, from lemma \ref{KernRate}, we derive that, for any $x$,\\ $\sup_{a\in\R^d_*}|f_{a,n}(\transp ax)-f_a(\transp ax)|=O_{\PP}(n^{-\frac{2}{4+d}})$. Then, let us consider $\Psi_j=\frac{f_{\check{a_j},n}(\transp{\check{a_j}}x)}{\check g^{(j-1)}_{\check{a_j},n}(\transp{\check{a_j}}x)}-\frac{f_{a_j}(\transp{a_j}x)}{g^{(j-1)}_{a_j}(\transp{a_j}x)}$, we have 
$\Psi_j=\frac{1} 
{\check g^{(j-1)}_{\check{a_j},n}(\transp{\check{a_j}}x)g^{(j-1)}_{a_j}(\transp{a_j}x)}$
$((f_{\check{a_j},n} (\transp{\check{a_j}}x)-f_{a_j}(\transp{a_j}x))g^{(j-1)}_{a_j}(\transp{a_j}x)+f_{a_j}(\transp{a_j}x)(g^{(j-1)}_{a_j}(\transp{a_j}x)-\check g^{(j-1)}_{\check{a_j},n}(\transp{\check{a_j}}x)))$,\\ i.e. $|\Psi_j|=O_{\PP}(n^{-\frac{1}{2}{\bf 1}_{d=1}-\frac{2}{4+d}{\bf 1}_{d>1}})$ since $f_{a_j}(\transp{a_j}x)=O(1)$ and $g^{(j-1)}_{a_j}(\transp{a_j}x)=O(1)$. We can therefore conclude similarly as in theorem \ref{Superdiffconv}.
\hfill$\Box$\\
\noindent{\bf Proof of theorem \ref{Kernelfiloiestimateurs}.}
We get ththeorem through theorem \ref{filoiestimateurs} and proposition \ref{QuotientDonneLoi}.\hfill$\Box$\\
\noindent{\bf Proof of theorem \ref{filoiestimateurs}.}
First of all, let us remark that hypotheses $(H1)$ to $(H3)$ imply that $\check \gamma_n$ and $\check c_n(a_k)$ converge towards $a_k$ in probability.\\
Hypothesis $(H4)$ enables us to derive under the integrable sign after calculation,\\
$\PP\frac{\dr}{\dr b}M(a_k,a_k)=\PP\frac{\dr}{\dr a}M(a_k,a_k)=0,$\\
$\PP\frac{\dr^2}{\dr a_i\dr b_j}M(a_k,a_k)=\PP\frac{\dr^2}{\dr b_j\dr a_i}M(a_k,a_k)=\int \varphi"(\frac{gf_{a_k}}{fg_{a_k}}) \frac{\dr}{\dr a_i}\frac{gf_{a_k}}{fg_{a_k}}\frac{\dr}{\dr b_j}\frac{gf_{a_k}}{fg_{a_k}}\ f\ dx,$\\
$\PP\frac{\dr^2}{\dr a_i\dr a_j}M(a_k,a_k)=\int \varphi'(\frac{gf_{a_k}}{fg_{a_k}})\frac{\dr^2}{\dr a_i\dr a_j}\frac{gf_{a_k}}{fg_{a_k}}\ f\ dx,$\\
$\PP\frac{\dr^2}{\dr b_i\dr b_j}M(a_k,a_k)=-\int \varphi"(\frac{gf_{a_k}}{fg_{a_k}}) \frac{\dr}{\dr b_i}\frac{gf_{a_k}}{fg_{a_k}}\frac{\dr}{\dr b_j}\frac{gf_{a_k}}{fg_{a_k}}\ f\ dx$,\\ and consequently 
$\PP\frac{\dr^2}{\dr b_i\dr b_j}M(a_k,a_k)=-\PP\frac{\dr^2}{\dr a_i\dr b_j}M(a_k,a_k)=-\PP\frac{\dr^2}{\dr b_j\dr a_i}M(a_k,a_k),$
which implies,
$\frac{\dr^2}{\dr a_i\dr a_j}K(g\frac{f_{a_k}}{g_{a_k}},f)=\PP\frac{\dr^2}{\dr a_i\dr a_j}M(a_k,a_k)-\PP\frac{\dr^2}{\dr b_i\dr b_j}M(a_k,a_k),$

$=\PP\frac{\dr^2}{\dr a_i\dr a_j}M(a_k,a_k)+\PP\frac{\dr^2}{\dr a_i\dr b_j}M(a_k,a_k),$
$=\PP\frac{\dr^2}{\dr a_i\dr a_j}M(a_k,a_k)+\PP\frac{\dr^2}{\dr b_j\dr a_i}M(a_k,a_k).$\\
The very definition of the estimators $\check \gamma_n$ and $\check c_n(a_k)$, implies that
$
\left\{
\begin{array}{rl}
\Pn_n\frac{\dr}{\dr b}M(b,a)=0\\
\Pn_n\frac{\dr}{\dr a}M(b(a),a)=0
\end{array}
\right.
$\\
ie $\left\{
\begin{array}{rl}
\Pn_n\frac{\dr}{\dr b}M(\check c_n(a_k),\check \gamma_n)=0\\
\Pn_n\frac{\dr}{\dr a}M(\check c_n(a_k),\check \gamma_n)+\Pn_n\frac{\dr}{\dr b}M(\check c_n(a_k),\check \gamma_n)\frac{\dr}{\dr a}\check c_n(a_k)=0,
\end{array}
\right.$ i.e.
$\left\{
\begin{array}{rl}
\Pn_n\frac{\dr}{\dr b}M(\check c_n(a_k),\check \gamma_n)=0\ (E0)\\
\Pn_n\frac{\dr}{\dr a}M(\check c_n(a_k),\check \gamma_n)=0\ (E1)
\end{array}
\right.$.\\
Under $(H5)$ and $(H6)$, and using a Taylor development of the  $(E0)$ (resp. $(E1)$) equation, we infer there exists $(\overline c_n, \overline \gamma_n)$ (resp. $(\tilde c_n, \tilde \gamma_n)$) on the interval $[(\check c_n(a_k),\check \gamma_n),(a_k,a_k)]$ such that \\
$-\Pn_n\frac{\dr}{\dr b}M(a_k,a_k)=[\transp{(\PP\frac{\dr^2}{\dr b\dr b}M(a_k,a_k))}+o_{\PP}(1),\transp{(\PP\frac{\dr^2}{\dr a\dr b}M(a_k,a_k))}+o_{\PP}(1)]a_n.$\\
(resp. $-\Pn_n\frac{\dr}{\dr a}M(a_k,a_k)=[\transp{(\PP\frac{\dr^2}{\dr b\dr a}M(a_k,a_k))}+o_{\PP}(1),\transp{(\PP\frac{\dr^2}{\dr a^2}M(a_k,a_k))}+o_{\PP}(1)]a_n$)\\
with $a_n=(\transp{(\check c_n(a_k)-a_k)},\transp{(\check \gamma_n-a_k)})$.
Thus we get
\begin{eqnarray}
\sqrt n a_n&=&\sqrt n
\left[
\begin{array}{ccc}
\PP\frac{\dr^2}{\dr b^2}M(a_k,a_k) & \PP\frac{\dr^2}{\dr a\dr b}M(a_k,a_k)\\
\PP\frac{\dr^2}{\dr b\dr a}M(a_k,a_k) & \PP\frac{\dr^2}{\dr a^2}M(a_k,a_k) \\
\end{array}
\right]^{-1}
\left[
\begin{array}{ccc}
-\Pn_n\frac{\dr}{\dr b}M(a_k,a_k)\\
-\Pn_n\frac{\dr}{\dr a}M(a_k,a_k)\\
\end{array}
\right]+o_{\PP}(1)\nonumber
\end{eqnarray}

$=\sqrt n(\PP\frac{\dr^2}{\dr b\dr b}M(a_k,a_k)\frac{\dr^2}{\dr a\dr a}K(g\frac{f_{a_k}}{g_{a_k}},f))^{-1}$\\

$.\left[
\begin{array}{ccc}
\PP\frac{\dr^2}{\dr b\dr b}M(a_k,a_k)+\frac{\dr^2}{\dr a\dr a}K(g\frac{f_{a_k}}{g_{a_k}},f) & \PP\frac{\dr^2}{\dr b\dr b}M(a_k,a_k)\\
\PP\frac{\dr^2}{\dr b\dr b}M(a_k,a_k) &  \PP\frac{\dr^2}{\dr b\dr b}M(a_k,a_k)\\
\end{array}
\right].\left[
\begin{array}{ccc}
-\Pn_n\frac{\dr}{\dr b}M(a_k,a_k)\\
-\Pn_n\frac{\dr}{\dr a}M(a_k,a_k)\\
\end{array}
\right]+o_{\PP}(1)$\\
Moreover, the central limit theorem implies: 
$\Pn_n\frac{\dr}{\dr b}M(a_k,a_k)\cvL \cN_d(0,\PP\|\frac{\dr}{\dr b}M(a_k,a_k)\|^2)$,\\
$\Pn_n\frac{\dr}{\dr a}M(a_k,a_k)\cvL \cN_d(0,\PP\|\frac{\dr}{\dr a}M(a_k,a_k)\|^2)$,
since $\PP\frac{\dr}{\dr b}M(a_k,a_k)=\PP\frac{\dr}{\dr a}M(a_k,a_k)=0$, which leads us to the result.\hfill$\Box$\\
\noindent{\bf Proof of proposition \ref{cvl}.}
Let us consider $\psi$ (resp. $\psi^{(k)}$) the characteristic function of  $f$ (resp. $g^{(k-1)}$). Let also consider the sequence $(a_i)$ defined in (\ref{VraiDefOfAK}) page \pageref{VraiDefOfAK}. \\We have
$|\psi(t)-\psi^{(k)}(t)|\leq \int |f(x)-g^{(k)}(x)|dx\leq K(g^{(k)},f).$ As explained in section 14 of Huber's article and through remark \ref{criteria-H} page \pageref{criteria-H} as well as through the additive relation of proposition \ref{lemmeHuber0prop}, we can say that $\lim_kK(g^{(k-1)}\frac{f_{a_k}}{[g^{(k-1)}]_{a_k}},f)=0$.
Consequently, we get $\lim_k g^{(k)}=f$.\\
\noindent{\bf Proof of theorem \ref{limnk}.}
We recall that $g_n^{(k)}$ is the kernel estimator of $\check g^{(k)}$.
Since the relative entropy is greater than the $L^1$-distance, we then have

$\lim_n\lim_k K(g_n^{(k)},f_n)\geq \lim_n\lim_k \int |g_n^{(k)}(x)-f_n(x)|dx$\\
Moreover, the Fatou's lemma implies that

$\lim_k \int |g_n^{(k)}(x)-f_n(x)|dx\geq \int \lim_k \big[|g_n^{(k)}(x)-f_n(x)|\big]dx=\int |[\lim_k g_n^{(k)}(x)]-f_n(x)|dx$

\noindent and
$\lim_n \int |[\lim_k g_n^{(k)}(x)]-f_n(x)|dx\geq \int \lim_n \big[|[\lim_k g_n^{(k)}(x)]-f_n(x)|\big]dx$

$\ \ \ \ \ \ \ \ \ \ \ \ \ \ \ \ \ \ \ \ \ \ \ \ \ \ \ \ \ \ \ \ \ \ \ \ \ \ \ \ $ $=\int |[\lim_n\lim_k g_n^{(k)}(x)]-\lim_nf_n(x)|dx$.\\
Trough lemma \ref{FromSection14Huber}, we then obtain that $0=\lim_n\lim_k K(g_n^{(k)},f_n)\geq \int |[\lim_n\lim_k g_n^{(k)}(x)]-\lim_nf_n(x)|dx\geq0$, i.e. that $\int |[\lim_n\lim_k g_n^{(k)}(x)]-\lim_nf_n(x)|dx=0$.\\
Moreover, for any given $k$ and any given $n$, the function $g_n^{(k)}$ is a convex combination of multivariate Gaussian distributions. As derived at remark  \ref{implyEstimBounded} of page \pageref{implyEstimBounded}, for all $k$, the determinant of the covariance of the random vector - with density $g^{(k)}$ - is greater than or equal to the product of a positive constant times the determinant of the covariance of the random vector  with density $f$.
The form of the kernel estimate therefore implies that there exists an integrable function $\varphi$ such that, for any given $k$ and any given $n$, we have $|g_n^{(k)}|\leq \varphi$. \\
Finally, the dominated convergence theorem enables us to say that 
$\lim_n\lim_k g_n^{(k)}=\lim_n f_n=f$, since $f_n$ converges towards $f$ and since $\int |[\lim_n\lim_k g_n^{(k)}(x)]-\lim_nf_n(x)|dx=0$.\hfill$\Box$\\
\noindent{\bf Proof of theorem \ref{LOIDUCRITERE}.}
Through a Taylor development of $\Pn_nM(\check c_n(a_k),\check \gamma_n)$ of rank 2, we get at point $(a_k,a_k)$:\\
$\Pn_nM(\check c_n(a_k),\check \gamma_n)$
$=\Pn_nM(a_k,a_k)+\Pn_n\frac{\dr}{\dr a}M(a_k,a_k)\transp {(\check \gamma_n-a_k)}+\Pn_n\frac{\dr}{\dr b}M(a_k,a_k)\transp {(\check c_n(a_k)-a_k)}$

$+\frac{1}{2}\{\transp {(\check \gamma_n-a_k)}\Pn_n\frac{\dr^2}{\dr a\dr a}M(a_k,a_k)(\check \gamma_n-a_k)+\transp {(\check c_n(a_k)-a_k)}\Pn_n\frac{\dr^2}{\dr b\dr a}M(a_k,a_k)(\check \gamma_n-a_k)$

$+\transp {(\check \gamma_n-a_k)}\Pn_n\frac{\dr^2}{\dr a\dr b}M(a_k,a_k)(\check c_n(a_k)-a_k)+\transp {(\check c_n(a_k)-a_k)}\Pn_n\frac{\dr^2}{\dr b\dr b}M(a_k,a_k)(\check c_n(a_k)-a_k)\}$\\
The lemma below enables us to conclude.
\begin{lemme} 
Let $H$ be an integrable function and let $C=\int\ H\ d\PP$ and $C_n=\int\ H\ d\Pn_n$,

$\ \ \ \ \ \ \ \ \ \ \ \ $then, $C_n-C=O_{\PP}(\frac{1}{\sqrt n}).$
\end{lemme}
Thus we get $\Pn_nM(\check c_n(a_k),\check \gamma_n)=\Pn_nM(a_k,a_k)+O_{\PP}(\frac{1}{n}),$
\\ i.e. $\sqrt n(\Pn_nM(\check c_n(a_k),\check \gamma_n)-\PP M(a_k,a_k))=\sqrt n(\Pn_nM(a_k,a_k)-\PP M(a_k,a_k))+o_{\PP}(1).$ \\Hence $\sqrt n(\Pn_nM(\check c_n(a_k),\check \gamma_n)-\PP M(a_k,a_k))$ abides by the same limit distribution as\\ $\sqrt n(\Pn_nM(a_k,a_k)-\PP M(a_k,a_k))$, which is $\cN(0,Var_{\PP}(M(a_k,a_k)))$.\hfill$\Box$\\
\noindent{\bf Proof of theorem \ref{KernelLOIDUCRITERE}.}
Through proposition \ref{QuotientDonneLoi} and theorem \ref{LOIDUCRITERE}, we derive theorem \ref{KernelLOIDUCRITERE}..\hfill$\Box$\\
\noindent{\bf Proof of theorem \ref{Sum=Product}.}
We immediately get the proof from theorem \ref{limnk}.\hfill$\Box$\\
\noindent{\bf Proof of theorem \ref{FirstReg}.}
Since $\Phi(g^{(1)},f)=0$, then, through lemma \ref{OrthoOfVect}, we deduct that the density of $\transp {b_2}X/\transp {a_1}X$, with $a_1=(0,1)'$ and $b_2=(1,0)'$, is the same as the one of $\transp {b_2}Y/\transp {a_1}Y$.\\
Hence, we derive that $E(X_1/X_2)=E(Y_1/Y_2)$ and also that the regression between $X_1$ and $X_2$ is $X_1=E(Y_1)+\frac{Cov(Y_1,Y_2)}{Var(Y_2)}(Y_2-E(Y_2))+\varepsilon,$
where $\varepsilon$ is a centered random variable such that it is orthogonal to $E(X_1/X_2)$.\hfill$\Box$\\
\noindent{\bf Proof of theorem \ref{SecondReg}.}
We infer this proof similarly to the proof of theorem \ref{FirstReg} section.\hfill$\Box$\\
\noindent{\bf Proof of corollary \ref{SecondRegCoro}.}
Assuming first that the $b_k$ and the $a_i$ are the canonical basis of $\R^d$. Then, for any $i\not= j$, $Y_i$ is independent from $Y_j$, i.e. $E(Y_k/Y_1,...,Y_j)=E(Y_k)$. Consequently, the regression between $X_k$ and $(X_1,...,X_j)$ is given by $X_k=E(Y_k)+\varepsilon_k$ where $\varepsilon$ is a centered random variable such that it is orthogonal to $E(X_k/X_1,...,X_j)$.\\
At present, we derive the general case thanks to the methodology used in the proof of lemma \ref{ChangBasis} section with the transformation matrix $B=(a_1,...,a_j,b_{j+1},...,b_d)$.\hfill$\Box$

\end{document}